\documentclass[useAMS,usenatbib]{mn2e}
\usepackage{times}
\usepackage{amsmath}
\usepackage{graphicx} 
\usepackage{multicol}
\usepackage{amssymb}
\usepackage{upgreek}
\usepackage{bbm}
\usepackage{dsfont}
\usepackage{sansmath}
\def\d{{\rm d}}
\def\D{{\rm D}}

\def\bfOmega{\mbox{\boldmath{$\Omega$}}}
\def\bfomega{\mbox{\boldmath{$\upomega$}}}

\def\bfPsi{\mbox{\boldmath{$\Psi$}}}

\def\matrixbf#1{\mbox{\large\textbf{\textsf{#1}}}}
\def\LD{L_{\rm{\tiny D}}}
\def\EKE{\,{\it EKE\ }}
\def\APE{\,{\it APE\ }}
\usepackage{soul}
\usepackage{color}
\definecolor{com_red}{rgb}{.8,.2,0.1}
\definecolor{ins_green}{rgb}{.1,.5,0.1}

\title[Baroclinic Instability on Hot Extrasolar Planets]{Baroclinic
  Instability on Hot Extrasolar Planets}

\author [I. Polichtchouk and
J. Y-K. Cho]{I. Polichtchouk\thanks{E-mail: I.Polichtchouk@qmul.ac.uk;
    J.Cho@qmul.ac.uk}\thanks{Corresponding author} and
  J. Y-K. Cho\footnotemark[1]\\
  Astronomy Unit, School of Physics and Astronomy, Queen Mary
  University of London, London E1 4NS, UK}

\begin{document}

\date{Accepted 2012 May 15. Received 2012 March 21; in original form
  2011 December 31}

\pagerange{\pageref{firstpage}--\pageref{lastpage}} \pubyear{2012}

\maketitle

\label{firstpage}

\begin{abstract}
  We investigate baroclinic instability in flow conditions relevant to
  hot extrasolar planets.  The instability is important for
  transporting and mixing heat, as well as for influencing large-scale
  variability on the planets.  Both linear normal mode analysis and
  non-linear initial value calculations are carried out -- focusing on
  the freely-evolving, adiabatic situation.  Using a high-resolution
  general circulation model (GCM) which solves the traditional
  primitive equations, we show that large-scale jets similar to those
  observed in current GCM simulations of hot extrasolar giant planets
  are likely to be baroclinically unstable on a timescale of few to
  few tens of planetary rotations, generating cyclones and
  anticyclones that drive weather systems.  The growth rate and scale
  of the most unstable mode obtained in the linear analysis are in
  qualitative, good agreement with the full non-linear calculations.
  In general, unstable jets evolve differently depending on their
  signs (eastward or westward), due to the change in sign of the jet
  curvature.  For jets located at or near the equator, instability is
  strong at the flanks -- but not at the core.  Crucially, the
  instability is either poorly or not at all captured in simulations
  with low resolution and/or high artificial viscosity.  Hence, the
  instability has not been observed or emphasized in past circulation
  studies of hot extrasolar planets.
\end{abstract}

\begin{keywords}
  hydrodynamics -- planets and satellites: atmospheres -- methods:
  numerical -- instabilities -- turbulence -- waves
\end{keywords}

\section{Introduction}\label{introduction}

Baroclinic instability is a generic flow instability that occurs in
rotating, stably-stratified fluids subject to a meridional (northward)
temperature gradient.  Examples of such a fluid are planetary
atmospheres and oceans.  The temperature gradient induces a vertical
(altitudinal) shear in the mean flow by thermal wind balance
\citep[e.g.][]{Pedlosky}; hence, baroclinic flows are those that
nominally vary in the vertical direction.  The instability itself is
important because it gives rise to large- and meso-scale weather
systems on planets.  It also serves as a source of turbulence, which
has been invoked as the initial condition in some simulations of
extrasolar planets to generate plausible initial jet profiles
\citep[e.g.][]{Cho03,Cho08a}.  More importantly, the instability is a
source of spatio-temporal variability which could be observed
remotely.

Baroclinic instability on extrasolar planets has not been studied thus
far.  In this work we perform a simple linear analysis of a
horizontally uniform jet in vertical shear.  We also use a
highly-accurate pseudospectral general circulation model (GCM) which
solves the hydrostatic primitive equations to study the non-linear
evolution of a non-uniform, gradient-wind balanced jet on an
extrasolar planet.  The primitive equations solved govern the
large-scale dynamics of planetary atmospheres (e.g. Holton 1992; see
also Cho et al.\ 2008 for some discussion relevant to the present
work).  Here the main focus is on close-in gaseous planets, as they
remain the best studied type of extrasolar planets thus far.  However,
much of the findings and discussion apply to hot extrasolar planets in
general -- regardless of whether a solid boundary is present or the
radiatively stable layer extends deeply into the planet.

For concreteness, we present calculations for a model planet with
physical parameters appropriate for the extrasolar giant planet
HD209458b (Table~1).  We focus on the stability of broad, high-speed
zonal jets -- positive (eastward) at the equator and negative
(westward) at high latitude -- under adiabatic (i.e. heating and
cooling rates balanced in the net) situation.  By `broad' we mean a
width of $\sim\!\LD$, where $\LD$ is the Rossby deformation length
(section~\ref{csp}), and `high-speed' means the speed is
$\sim$\,1000~m~s$^{-1}$ at the core of the jet.  Such jets are
commonly produced in diabatically-forced GCM simulations of close-in
extrasolar giant planets
\citep[e.g.][]{Showman08,Rauscher10,Thrastarson10}.  A study of
adiabatic behaviour is needed because it provides the necessary
baseline for comparing the instability under forced conditions and
because, in many circumstances, the produced jets are not maintained
by the applied thermal forcing (but some flow-modified version, away
from the specified radiative equilibrium).

Our basic approach in this work is to carefully study baroclinic
instability in sufficient generality, without complicating the
fundamental process with details which are still uncertain for
extrasolar planets.  The primary aim here is three-fold: 1) to
ascertain the importance of baroclinic instability as a generic
process operating on extrasolar planets; 2) to gain a better
understanding of the outputs from current extrasolar planet GCM
simulations, made difficult by the complexity of solving the primitive
equations accurately; and, 3) to explore fundamental issues in
baroclinic instability that have received less emphasis in traditional
geophysical fluid dynamics studies, due to the markedly different
parameter regime of many extrasolar planets compared to that of the
Earth.

The overall plan of the paper is as follows.  Section~\ref{linear}
presents linear stability analysis.  Linear growth rates and phase
speeds are calculated for the traditional primitive equations on the
`$\beta$-plane', a differentially rotating plane tangent to the
surface of the planet at a given latitude.  In
section~\ref{non-linear} we present the non-linear evolution of the
instability, obtained from full numerical simulations.  This section
also presents the description of the numerical model and setup, as
well as the non-convergence of under-resolved and/or over-dissipated
simulations.  The foundation for baroclinic life-cycle study is also
laid in this section; a detailed discussion of the phenomenon is
presented elsewhere, as are of forced evolution and transient growth.
Recapitulation and discussion are given in section~\ref{summary}.

\section{Linear Theory}\label{linear}

\subsection{Charney-Stern-Pedlosky Criteria}\label{csp}

Necessary conditions for instability exist.  These may be derived
directly from global conservation of pseudoenergy and are given in
\citet{Charney62} and \citet{Pedlosky64}.  Hence, we shall not derive
the conditions here but simply list them for the reader's convenience.
The conditions play an important role in this work, particularly in
understanding the setup of the nonlinear initial value problem
(section~\ref{non-linear}).

Let $q = \alpha\,(\bfomega + 2\bfOmega)\cdot\nabla^{\mbox{\tiny{\sc
      L}}}\varphi$ be the potential vorticity, where $\bfomega$ is the
relative vorticity, $\bfOmega$ is the planetary vorticity, $\alpha$ is
the specific volume ($ = 1/\rho$, where $\rho$ is the density) and
$\nabla^{\mbox{\tiny{\sc L}}}$ is a gradient operator acting on a
materially conserved field $\varphi$, which may be a function of
temperature and pressure (e.g.  potential temperature or entropy).
Additionally, let $x$, $y$ and $z$ be the zonal~(${\bf i}$),
meridional~(${\bf j}$) and vertical (${\bf k}$) directions,
respectively.  Given the zonal flow, ${\bf U} = U(y,z)\,{\bf i}$, and
the basic state potential vorticity $Q(y,z)$ such that $q(x,y,z) = Q +
q^\prime(x,y,z)$, one of the following necessary criteria must be met
for the onset of instability:
\begin{enumerate}

\item \ \ \ $\partial Q / \partial y$ and $\partial U / \partial z$
  are opposite signs at the upper \\ \hspace*{1cm} boundary\\*[-1mm]

\item \ \ $\partial Q / \partial y$ and $\partial U / \partial z$
  are same signs at the lower\\ \hspace*{1cm} boundary\\*[-1mm]

\item \ $\partial U / \partial z$ is the same sign at the upper {\it
    and} lower\\ \hspace*{1cm} boundaries -- a condition that is
  distinct from\\ \hspace*{1cm} condition (i) or (ii), if $\partial Q
  / \partial y = 0$\\*[-1mm]

\item \ $\partial Q / \partial y$ changes sign somewhere in the
  interior.

\end{enumerate}
Note that $Q = Q({\bf U})$ and the prime denotes deviation from the
basic state.

\begin{table}
  \centering
  \caption{Numerical parameter values for HD209458b.  Here $g$ 
    is surface gravity; $R_p$ is equatorial radius; $\Omega$ is 
    rotation rate; ${\cal R}$ is gas constant; $c_p$ is specific 
    heat at constant pressure; $H$ is scale height; $p_{\rm r}$ 
    is reference pressure; $T_{\rm eq}$ is equilibrium temperature; 
    $U$ is characteristic flow speed; and, $N$ is 
    Brunt-V\"{a}is\"{a}l\"{a} frequency.}
    \label{table-params}
    \vspace*{.2cm}
    \begin{tabular}{lll} \hline \hline
      Parameter \hspace*{3mm} & Value \hspace*{1cm} & Units \\
      \hline
      $g$   & 9.8  &  m~s$^{-2}$ \\
      $R_{p}$   &  $10^{8}$     & m \\
      $\Omega$   & $2.1\!\times\! 10^{-5}$    &  s$^{-1}$ \\
      ${\cal R}$  & $3.5\!\times\! 10^3$     & J~kg$^{-1}$~K$^{-1}$  \\
      $c_{p}$   & $1.23\!\times\! 10^4$     &   J~kg$^{-1}$K$^{-1}$   \\
      $H$   & $5.8\!\times\! 10^5$     &  m  \\
      $p_{\rm r}$   & $10^5$     &  Pa   \\
      $T_{\rm eq}$   & 1500     &  K   \\
      $U$   & 1000    &  m~s$^{-1}$   \\
      $N$   & $2\!\times\! 10^{-3}$     &  s$^{-1}$   \\
      \hline\hline 
    \end{tabular}
\end{table}

For realistic flow profiles studied in section~\ref{non-linear}, the
instability criterion is normally satisfied through criterion (iv).
In addition, criteria (i) and (ii) are also satisfied in most
cases. These conditions are useful for assessing stability of any
basic flow configuration.  However, they do not provide quantitative
information, such as the growth rates of unstable modes and phase
speeds of waves/eddies generated by the instability.

For the Earth, the stability analysis is typically based on the
quasi-geostrophic (QG) theory, in which small Rossby number $Ro$ and
order unity Burger number $Bu$ are assumed
\citep{Charney47,Eady49,Phillips51}.  Given the characteristic flow
speed $U$, Coriolis parameter~$f$, horizontal length scale $L$, and
the Rossby deformation length scale $\LD$, $Ro$ and $Bu$ are defined
$Ro\! =\! U/(fL)$ and $Bu\! =\! (\LD/L)^2$, respectively.  Here $\LD\!
=\! NH/f$ is the Rossby deformation length, where $N$ is the
Brunt-V\"{a}is\"{a}l\"{a} frequency; $H$ is the characteristic
vertical scale; and, $f = 2\Omega\sin\phi$, where $\Omega =
|\bfOmega|$ is the planetary rotation rate and $\phi$ is the latitude.
Note that both $Ro$ and $Bu$ vary with latitude.  For example,
formally, $Ro \rightarrow\infty$ as $\phi\rightarrow 0$.

In QG theory, adiabatic dynamics is governed by the material advection
of potential vorticity:
\[    
\frac{\D q_{\mbox{\tiny{\sc QG}}}}{\D t}\ =\ 0\, ,
\]
where $\D / \D t$ is the material derivative and
\[
q_{\mbox{\tiny{\sc QG}}}\ =\ f + \nabla^2\psi + \alpha
f_0^2 \frac{\partial}{\partial z}\left(\frac{1}{\alpha N^2}
  \frac{\partial\psi}{\partial z} \right)
\]
is the QG potential vorticity in the `$\beta$-plane approximation'
(see section~\ref{analysis}).  Here $f = f_0 + \beta y$, where $f_0 =
f(\phi_0)$ and $\beta = (\d f/\d y)|_{\phi = \phi_0}$ for a specific
latitude $\phi_0$; $\psi$ is the streamfunction; and, $\nabla^2$ is
the horizontal Laplacian operator.  Note that $q_{\mbox{\tiny{\sc
      QG}}}$ can be inverted -- as with the full primitive equation
$q$, under the QG balance condition -- to obtain all other dynamical
variables \citep{Hoskins}.

The QG equations (the above advection equation for $q_{\mbox{\tiny{\sc
      QG}}}$ plus boundary conditions) derive from the more complete
primitive equations \citep[e.g.][]{Pedlosky}.  The standard QG
equations are appropriate for large-scale motions on many planets,
away from the low latitudes.  However, they are not
broadly\footnote{QG theory may still be valid locally on hot
  extrasolar planets.}  appropriate for a large number of extrasolar
planets, which are characterized by $Ro$ of order unity (see
Table~\ref{table-params}) -- even away from the equatorial region.
More importantly, much dynamics of interest on extrasolar planets
occur in the equatorial region (section~\ref{equatorial}), where the
traditional QG theory does formally break down.  Therefore, we perform
our stability analysis using the full primitive equations.

\subsection{Governing Equations} 

In the standard pressure ($p$) coordinate system \citep{Kasahara74},
the hydrostatic primitive equations read:
\begin{subequations}\label{PE}
 \begin{eqnarray}
   \lefteqn{\frac{\partial{\bf v}}{\partial t} + 
     {\bf{v}}\!\cdot\!\tilde\nabla\,{\bf v} +
     \omega\frac{\partial{\bf v}}{\partial p} + 
     f{\bf {k}}\times {\bf{v}} + \tilde\nabla \Phi\ =\ 
     {\cal F}_{{\bf{v}}}}\\
   \lefteqn{\frac{\partial \theta}{\partial t} + 
     {\bf v}\!\cdot\!\tilde\nabla\,\theta + 
     \omega\frac{\partial\theta}{\partial p}\ =\ {\cal F}_{\theta} }\\
   \lefteqn{\,\tilde\nabla\! \cdot\! {\bf{v}} + 
     \frac{\partial \omega}{\partial p}\ =\ 0}\\
   \lefteqn{\frac{\partial \Phi}{\partial p} + h\theta\ =\ 0\, .}
 \end{eqnarray}
\end{subequations}     
Here ${\bf{v}}({\bf{x}},t)= (u,v)$ is the (zonal,\,meridional)
velocity in the frame rotating with $\bfOmega$, where ${\bf
  x}\in\mathbb{R}^3$; $\omega = {\tilde\D}p/{\tilde\D}t$ is the
`vertical' velocity, where ${\tilde\D}/{\tilde\D} t
= \partial/\partial t + {\bf v}\!\cdot\!\tilde\nabla +
\omega\partial/\partial p$ with $\tilde\nabla$ operating along
constant surfaces of $p$ (which in general is not materially
conserved); $\theta = T(p_{\rm r}/p)^\kappa$ is the potential
temperature, where $T$ is the temperature, $p_{\rm r}$ is the
reference pressure, $\kappa = {\cal R}/c_p$ with ${\cal R}$ the gas
constant and $c_p$ the specific heat at constant pressure; $\Phi = gz$
is the geopotential above the planetary radius $R_p$, where $g$ is the
constant surface gravity; $h(p) = {\cal R\,}(p/p_{\rm
  r})^{\kappa}/\,p\,$; and, ${\cal F}_{\bf{v}}$ and ${\cal
  F}_{\theta}$ are, respectively, momentum and potential temperature
sources/dissipations.  From here on, we exclusively work in
$p$-coordinate and drop the tilde over the advective and gradient
operators for notational clarity.

The above set of equations is closed with the ideal gas law,
$p\,\alpha = {\cal R}T$.  The equations are also supplemented with the
boundary conditions,
\begin{equation}\label{BC}
  \omega = 0 \qquad {\rm at} \qquad  p = 0,\, p_{\rm r}\, .
\end{equation}
Hence, the domain boundaries are material surfaces and no mass flows
across them.

\subsection{Two-Layer, Beta-Plane Analysis}\label{analysis}

In this section, we linearize equations~(\ref{PE}) on the
$\beta$-plane, where $f(\phi)$ is represented by $f(y) = f_0 + \beta
y$.  Here, $f_0$ and $\beta$ are constants, $y = R_p (\phi - \phi_0)$
and the motion is assumed to be periodic in the zonal direction with
no meridional component at the latitudinal boundaries.  The
$\beta$-plane is a tangent plane located at $\phi_0$, and the setup is
only formally justified for scales that are small compared to $R_p$.
However, in practice the $\beta$-plane approximation mainly results in
small distortion of planetary waves and captures the essential
qualitative behavior.  For the analysis in this section, we neglect
source/dissipation terms in equations~(\ref{PE}) -- i.e. ${\cal
  F}_{\bf v}({\bf x},t) = {\cal F}_\theta({\bf x},t) = 0$.  This is
because, as discussed in section~\ref{introduction}, we are interested
in the dynamics of jets that result from conditions e.g. when
  the net heating is not large or the effective thermal relaxation
  time is not small.

\begin{figure} 
  \vspace*{.5cm}\hspace*{.1cm}
  \includegraphics[scale=0.4]{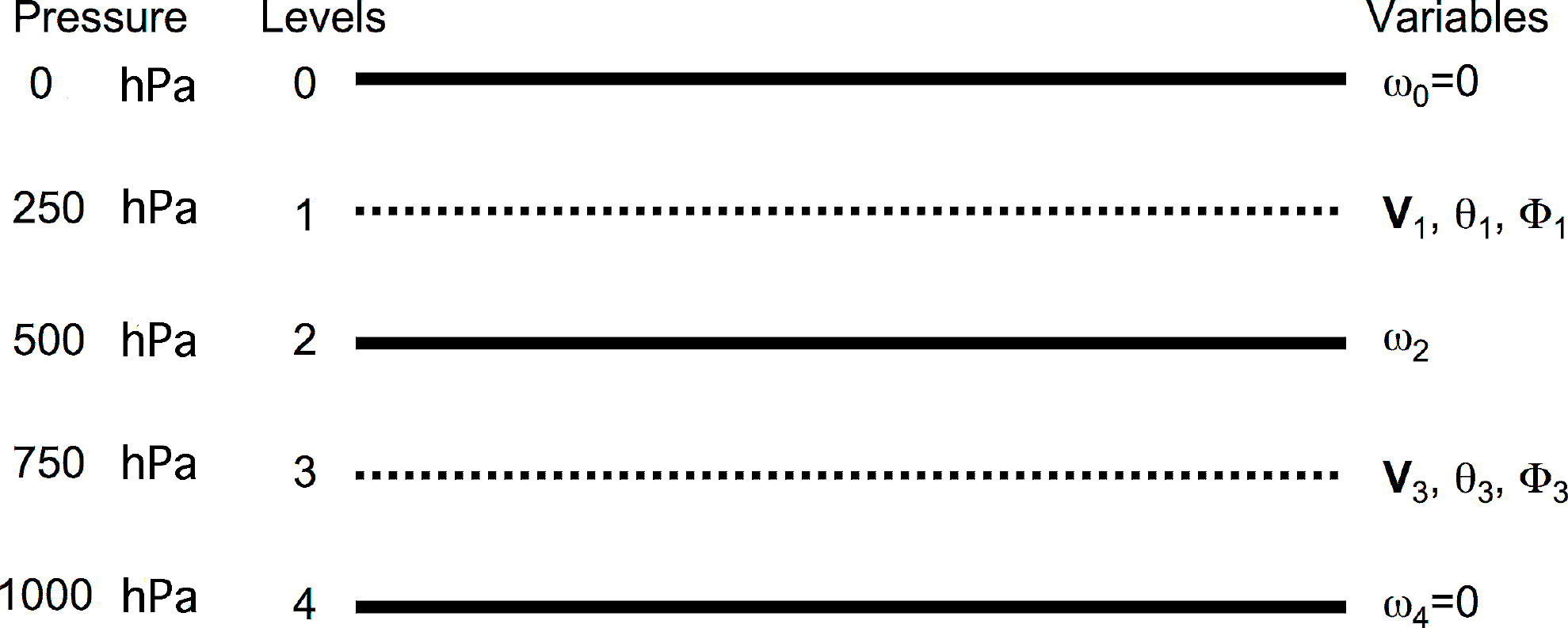}\vspace*{.3cm}
  \caption{Vertical structure of the two-layer primitive equations
    model.  Different field variables are defined on different levels;
    1~hPa = $10^2$~Pa = 1~mbar.  Bold lines are the layer boundaries.}  
  \label{fig1}
\end{figure}

We perform a standard normal mode analysis of the baroclinic
instability admitted by a two-layer representation of
equations~(\ref{PE}).  Similar work has been carried out by
\citet{Wiin-Nielsen} and \citet{Fraedrich} for the Earth.  As in these
studies, we simplify equation set (\ref{PE}) to that appropriate for a
discretised model with two equally-spaced, stacked layers in the
$p$-coordinate.  In this model ${\bf v}$, $\theta$ and $\Phi$ are
defined at odd levels and $\omega$ is defined at even levels.  The
structure is illustrated in Fig.~\ref{fig1}.  The equations for the
interior levels are:
\begin{subequations}\label{2L-PE}
\begin{eqnarray}
  \lefteqn{ \frac{\partial{\bf v}_1}{\partial t} + 
    {\bf v}_1\!\cdot\!\nabla {\bf v}_1 +  \omega_2\!
    \left(\frac{{\bf v}_3 - {\bf v}_1}{2\,\triangle p}\right) + 
    f{\bf k}\!\times\!{\bf v}_1\, =\, -\nabla \Phi_1 } \\
  \lefteqn{ \frac{\partial{\bf v}_3}{\partial t} + 
    {\bf v}_3\!\cdot\!\nabla{\bf v}_3 + \omega_2\!
    \left(\frac{{\bf v}_3 - {\bf v}_1}{2\,\triangle p}\right) + 
    f{\bf k}\!\times\!{\bf v}_3\, =\, -\nabla \Phi_3 } \\
  \lefteqn{\frac{\partial \theta_1}{\partial t} + \nabla\!\cdot\!
    (\theta_1\,{\bf v}_1) + \frac{\omega_2\theta_2}{\triangle p}\, 
    =\, 0 }\\
  \lefteqn{\frac{\partial \theta_3}{\partial t} + \nabla\!\cdot\!
    (\theta_3\,{\bf v}_3) - \frac{\omega_2\theta_2}{\triangle p}\, 
    =\, 0 }\\
  \lefteqn{\,\nabla\!\cdot\!{\bf v}_1 + \frac{\omega_2}{\triangle p}\,
    =\, 0}\\
  \lefteqn{\,\nabla\!\cdot\!{\bf v}_3 - \frac{\omega_2}{\triangle p}\, 
    =\, 0}\\
  \lefteqn{\,\Phi_1 - \Phi_3\, =\,  h_2\,\triangle p\,\theta_2\, ,}
\end{eqnarray}
\end{subequations} 
where $\triangle p = p_{\rm r}/2$ denotes the pressure difference
between odd or even numbered levels and $\theta_2 = \theta = (\theta_1
+ \theta_3) / 2$.  It follows from equations~(\ref{2L-PE}e) and
(\ref{2L-PE}f) that barotropic (vertically averaged) wind is
non-divergent.  In the present analysis, we take the
Brunt-V\"{a}is\"{a}l\"{a} frequency $N$ to be uniform; GCM simulations
by \citet{Thrastarson10} show static stability to be fairly constant
over one or two scale heights for a wide range of conditions.

Baroclinic instability in the two-level primitive equations system is
obtained from perturbations of an unstable `Eady-type' basic flow
with uniform vertical shear \citep{Eady49}:
\begin{eqnarray*}
  \bar{u}_1 & = & -\bar{u}_3\ =\  U_0 \\
  \bar{v}_1 & = & \ \ \bar{v}_3\ \, =\ 0 \\ 
  \bar{\omega}_1 & = & \ \ \bar{\omega}_3\  =\ 0 \\
  \bar{\theta}_1 & = & -\frac{2f_0}{h_2\,\triangle p }\, U_0\, y\,
  +\, \sigma_0 \\
  \bar{\theta}_3 & = & -\frac{2f_0}{h_2\,\triangle p }\, U_0\, y\, 
  -\, \sigma_0\, \\
  \bar{\theta}_2 & = & \ \ \bar{\theta}\ \ \ =\ 
  -\frac{2f_0}{h_2\,\triangle p }\, U_0\, y\, .
\end{eqnarray*}
Here $U_0\ (= U/2)$ characterizes the strength of the thermal wind and
its shear; and, $\sigma_0 = (\theta_1-\theta_3)/2$ is related to the
reference static stability, $S = - \alpha \ \partial \ln
\theta/\partial p$, through
\[
S\ =\ \frac{\sigma_0 h_2}{\triangle p}\, .
\]
For simplicity we shall consider meridionally-independent
perturbations applied to the above basic flow.  We now linearize
equations~(\ref{2L-PE}) about this basic state and arrive at the
following:
\begin{subequations}
\begin{eqnarray}
  \lefteqn{ \frac{\partial{\bf v}_1^\prime}{\partial t} + 
    U_0\,\frac{\partial{\bf v}_1^\prime}{\partial x} - 
    \frac{U_0}{\triangle p}\,\omega_2^\prime\,{\bf i} + 
    f{\bf k}\!\times\!{\bf v}_1^\prime\ =\ 
    - \frac{\partial\Phi_1^\prime}{\partial x} }\\
  \lefteqn{ \frac{\partial{\bf v}_3^\prime}{\partial t} - 
    U_0\,\frac{\partial{\bf v}_3^\prime}{\partial x} - 
    \frac{U_0}{\triangle p}\,\omega_2^\prime\,{\bf i} + 
    f{\bf k}\!\times\!{\bf v}_3^\prime\ =\ 
    - \frac{\partial\Phi_3^\prime}{\partial x} }\\
  \lefteqn{\frac{\partial \theta'}{\partial t} - 
    \frac{f_0\, U_0}{h_2\,\triangle p}(v_1^\prime + v_3^\prime) - 
    \sigma_0\frac{ \omega_2'}{\triangle p}\ =\ 0} \\
  \lefteqn{ \nabla\!\cdot\!{\bf v}_1^\prime + 
        \frac{\omega_2^\prime}{\triangle p}\ =\ 0 }\\
  \lefteqn{ \nabla\!\cdot\!{\bf v}_3^\prime - 
    \frac{\omega_2^\prime}{\triangle p}\ =\ 0 }\\
  \lefteqn{ \Phi_1^\prime - \Phi_3^\prime\ =\  
    h_2\,\triangle p\,\theta^\prime\, . }
\end{eqnarray}
\end{subequations}

The temperature equation~(4c) is obtained by summing
equations~(\ref{2L-PE}c) and (\ref{2L-PE}d) and linearising.  Further,
if we denote the vertical average of a variable $\xi$ by
\[
\xi_+ \equiv \frac{1}{2}(\xi_1 + \xi_3)
\] 
and the half vertical difference by 
\[
\xi_- \equiv \frac{1}{2}(\xi_1 - \xi_3)\, ,  
\]
summing and differencing equations~(4a) and (4b) give:
\begin{subequations}
\begin{eqnarray}
\lefteqn{ \frac{\partial{\bf v}_+^\prime}{\partial t} +
  U_0\,\frac{\partial{\bf v}_-^\prime}{\partial x}
  - \frac{U_0}{\triangle p}\omega_2^\prime\,{\bf i} +
  f{\bf k}\!\times\!{\bf v}_+^\prime\ =\ 
  -\frac{\partial\Phi_+^\prime}{\partial x} }\\
\lefteqn{ \frac{\partial{\bf v}_-^\prime}{\partial t} +
  U_0\,\frac{\partial{\bf v}_+^\prime}{\partial x} +
  f{\bf k}\!\times\!{\bf v}_-^\prime\ =\ 
  -\frac{\partial\Phi_-^\prime}{\partial x}. }
\end{eqnarray}
\end{subequations} 
By applying curl and divergence, we obtain vorticity and divergence
forms, respectively, of the above equations.  The equations set is
closed when potential temperature and pressure velocity are eliminated
using the hydrostatic and continuity equations.  We then introduce the
streamfunctions, $\psi_1$ and $\psi_3$, and the velocity potentials,
$\chi_1$ and $\chi_3$, for levels 1 and 3 such that
\[
\frac{\partial^2}{\partial x^2}\left(\chi_1 + \chi_3\right)\ =\ 0
\]
and obtain four evolution equations for the barotropic vorticity,
baroclinic vorticity, baroclinic divergence, and geopotential
(i.e. potential temperature):
\begin{eqnarray*}
  \frac{\partial^2\psi_+^\prime}{\partial x^2} & = &
  \frac{\partial^2}{\partial x^2}\!\left(\frac{\psi_1^\prime +
      \psi_3^\prime}{2}\right) \, , \\
  \frac{\partial^2\psi_-^\prime}{\partial x^2} & = &
  \frac{\partial^2}{\partial x^2}\!\left(\frac{\psi_1^\prime -
      \psi_3^\prime}{2}\right) \, , \\
  \frac{\partial^2\chi_-^\prime}{\partial x^2} & = &
  \frac{\partial^2}{\partial x^2}
  \left(\frac{\chi_1'-\chi_3^\prime}{2}\right)\ =\
  -\frac{\omega_2^\prime}{\triangle p} \, , \\
  \Phi_-^\prime\ \ & = & \frac{h_2\,\triangle p\,\theta^\prime}{2}\, ,
\end{eqnarray*}
respectively.  The evolution equations for these quantities are:
\begin{subequations}\label{Dimensional system}

\begin{eqnarray}
  \frac{\partial}{\partial t}\! \left( 
    \frac{\partial^2\psi_+^\prime}{\partial x^2} \right)\!\!\!
  & = & \!\!\! -U_0\frac{\partial}{\partial x}\!
  \left( \frac{\partial^2\psi_-^\prime}{\partial x^2} \right) - 
  \beta\frac{\partial\psi_+^\prime}{\partial x} \\
  \frac{\partial}{\partial t}\!\left(\frac{\partial^2
      \psi_-^\prime}{\partial x^2}\right)\!\!\!  & = & \!\!\!  
  -U_0\frac{\partial}{\partial x}\!
  \left( \frac{\partial^2 \psi_+^\prime}{\partial
      x^2} \right) - f_0\frac{\partial^2 \chi_-^\prime}{\partial
    x^2} - \beta\frac{\partial\psi_-^\prime}{\partial x} \\
  \frac{\partial}{\partial t}\!\left( \frac{\partial^2
      \chi_-^\prime}{\partial x^2} \right)\!\!\!  & = & \!\!\!   
  -\frac{\partial^2 \Phi_-^\prime}{\partial x^2} + f_0\frac{\partial^2
    \psi_-^\prime}{\partial x^2} - \beta
  \frac{\partial \chi_-^\prime}{\partial x} \\
  \frac{ \partial \Phi_-^\prime}{ \partial t}\ & = & \!  
  U_0 f_0 \frac{\partial\psi_+^\prime}{\partial x}
  -\frac{{\cal R}\, \sigma_0}{2^{\kappa+1}} 
  \frac{\partial^2\chi_-^\prime}{\partial x^2}\, . 
\end{eqnarray}
\end{subequations}
 
At this point, the foregoing system of equations can be made
non-dimensional for a more `generalized' treatment, as is typical in
instability studies.  However, we shall describe our analysis of the
equations presented in the dimensional form.  We feel this facilitates
a more lucid interpretation of the results in some ways.  For the
interested reader, we have included the non-dimensional account in
Appendix~A and refer the reader to that section, especially for the
dependence of the results on non-dimensional parameters.

Denoting disturbances by
\[
\bfPsi\ =\ \hat{\bfPsi}
\exp\{ik\,(x - c \, t)\}\, ,
\]
where $\bfPsi = (\psi_+' , \psi_-' , \chi_-' , \Phi_-')^{\rm T}$,
$\hat{\bfPsi} = (\hat{\Psi}_+ , \hat{\Psi}_- , \hat{\chi}_- ,
\hat{\Phi}_-)^{\rm T}$ and $c\in\mathbb{C}$,
equations~(\ref{Dimensional system}) reduce to
\[
\matrixbf{M}\,\hat\bfPsi\ =\ \mathbf{0}\, 
\]
with\\
\[
\matrixbf{M}\ =\ 
\begin{bmatrix} 
   -c - \beta/k^2 & U_0 & 0 & 0 \\ 
  U_0 & -c - \beta/k^2  & -i\, f_0/k & 0\\ 
  0 & i\, f_0/k & 
  -c - \beta/k^2 & -i/k \\
  f_0\, U_0 & 0 & -i\, k\, {\cal R}\, \sigma_0/2^{\kappa+1} & c
\end{bmatrix}
.
\]\\*[.2cm]
\noindent
For a non-trivial solution, $\mbox{det}(\matrixbf{M}) = 0$.  This
leads to a fourth-order characteristic equation for $c$\,:\\
\begin{eqnarray}\label{baroclinic_Dimfreq}
  \lefteqn{ c^4\, +\, 
    c^3\left(\frac{3\beta}{k^2}\right)\, +\,
    c^2\left(\frac{3\beta^2}{k^4}\, -\, 
      \frac{f_0^2}{k^2}-\frac{{\cal R} \, \sigma_0}{2^{\kappa+1}}-
      U_0^2\right)\, + } \nonumber \\
  \lefteqn{ c\left(\frac{\beta^3}{k^6}\, -\, 
      \frac{\beta \, f_0^2}{k^4}\, -\, 
      \frac{\beta \, {\cal R} \, \sigma_0}{2^{\kappa} \,k^2}\, -
      \frac{\beta \, U_0^2}{k^2}\right)\, +\, } \nonumber \\
  \lefteqn{ \left(\frac{{\cal R} \, \sigma_0 \,
        U_0^2}{2^{\kappa+1}}-\frac{f_0^2 \, U_0^2}{k^2}\, -\, 
    \frac{\beta^2 \, {\cal R} \, \sigma_0}{2^{\kappa+1} \,k^4}\right)\ \ 
  =\ \ 0\, .  }  
\end{eqnarray}\\*[-.4cm]

Equation~(\ref{baroclinic_Dimfreq}) is solved numerically for $c$ as a
function of $k$, while keeping the values of $f_0 \, ,\beta \, , U_0
\, ,{\cal R} \, ,\sigma_0 \,$ and $\kappa$ constant.  If
$\Im\mathfrak{m}\{c\}\ne 0$, the disturbances grow or decay
exponentially since they are proportional to $\exp\{-i\,k\,c\,t\}$.
Two of the roots of equation~(\ref{baroclinic_Dimfreq}) are stable
eastward- and westward-traveling inertia-gravity waves.  The other two
roots are baroclinic waves.  These waves propagate neutrally (i.e.
without growing or decaying) eastward and westward, if
$\Im\mathfrak{m}\{c\} = 0$ (provided $\Re\mathfrak{e}\{c\}\ne 0$).

\begin{figure*}
  \vspace*{0.3cm}
  \includegraphics[height=6.1cm, width=17.5cm]{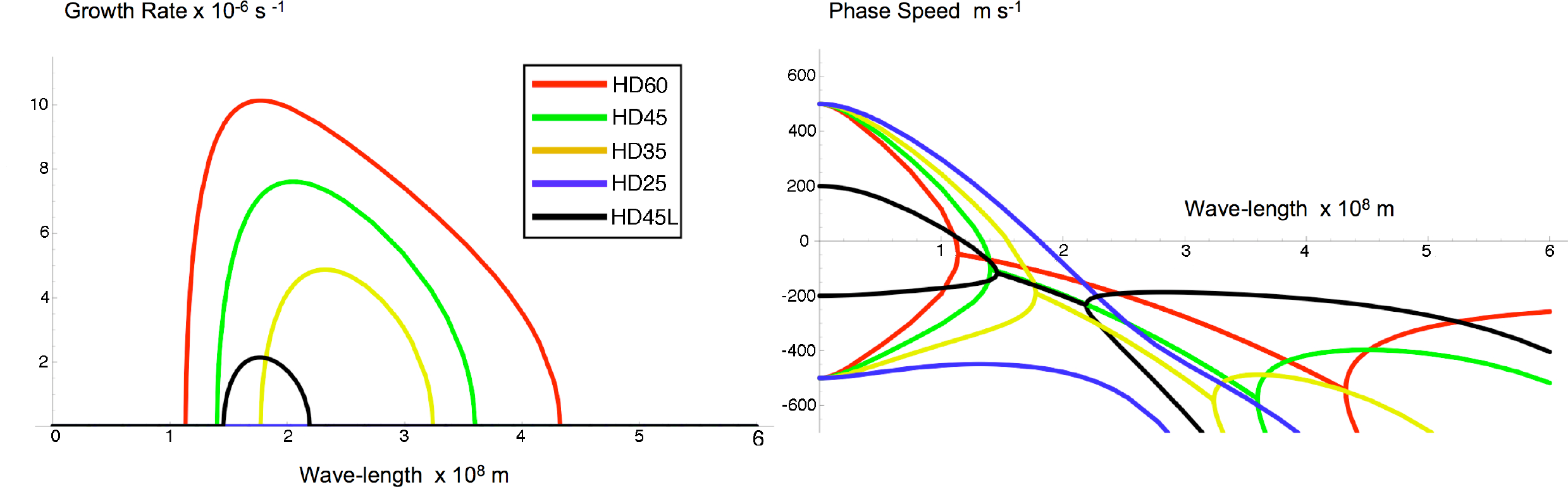}
  \vspace*{0.3cm}
  \caption{Growth rate [$k\cdot \Im\mathfrak{m}\{c\}$] (left) and
    phase speed [$\Re\mathfrak{e}\{c\}$] (right) for HD209458b, as a
    function of wavelength $2\pi\,k^{-1}$.  Curves `HD60', `HD45',
    `HD35' and `HD25' represent growth rates and phase speeds at $\phi
    = (60\degr, 45\degr, 35\degr, 25\degr)$; $f_0 = 4.2 \times
    10^{-5}\,\sin{\phi}$~s$^{-1}$, $\beta = 4.2 \times 10^{-13}\,
    \cos{\phi}$~m$^{-1}$~s$^{-1}$, $U_0 = 500$~m~s$^{-1}$, ${\cal R} =
    3500$~J~kg$^{-1}$~K$^{-1}$, $\sigma_0 = 300$~K and $\kappa =
    0.286$. Curve `HD45L' has been computed for HD209458b parameters
    at $\phi = 45\degr$, but with $U_0 = 200$~m~s$^{-1}$.}
  \label{fig2}
  \vspace*{0.3cm}
\end{figure*}

\begin{figure*}
  \vspace*{1cm}
  \includegraphics[height=6.1cm, width=17.5cm]{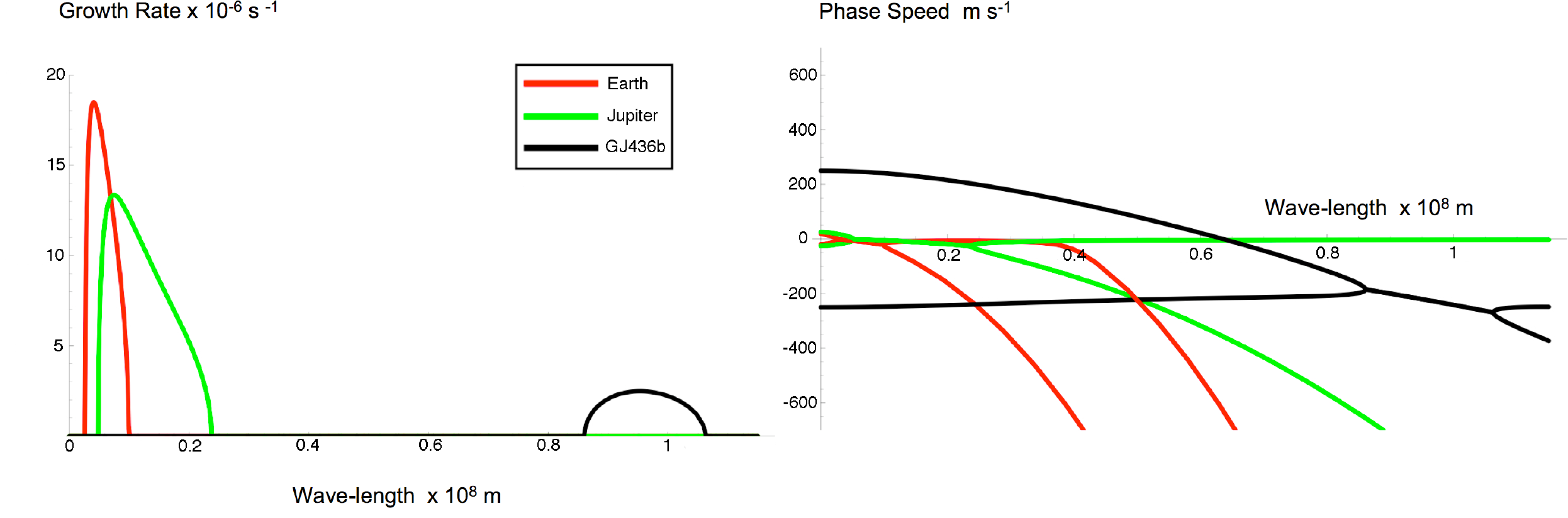}
  \vspace*{0.3cm}
  \caption{Growth rate [$k\cdot \Im\mathfrak{m}\{c\}$] (left) and
    phase speed [$\Re\mathfrak{e}\{c\}$] (right) at $\phi = 45\degr$
    for Earth, Jupiter and GJ436b as a function of wavelength $2
    \pi\,k^{-1}$.  For the Earth, $f_0 = 10^{-4}$~s$^{-1} $, $\beta =
    1.6 \times 10^{-11}$~m$^{-1}$~s$^{-1}$, $U_0 = 20$~m~s$^{-1}$,
    ${\cal R} = 287$~J~kg$^{-1}$~K$^{-1}$, $\sigma_0 = 15$~K, $\kappa
    = 0.286$.  For Jupiter, $f_0 = 2.5 \times 10^{-4}$~s$^{-1}$,
    $\beta = 3.5 \times 10^{-12}$~m$^{-1}~$s$^{-1}$, $U_0 =
    25$~m~s$^{-1}$, ${\cal R} = 3779$~J~kg$^{-1}$~K$^{-1}$, $\sigma_0
    = 24$~K, $\kappa = 0.286$.  For GJ436b, $f_0 = 3.9 \times
    10^{-5}$~s$^{-1}$, $\beta = 1.4 \times
    10^{-12}$~m$^{-1}$~s$^{-1}$, $U_0 = 250$~m~s$^{-1}$, ${\cal R} =
    3500$~J~kg$^{-1}$~K$^{-1}$, $\sigma_0 = 150$~K, $\kappa = 0.286$.
    Note, $\sigma_0$ values for Jupiter and GJ436b have been computed
    assuming constant temperatures of 120~K and 750~K, respectively.
    Note the change in scales, compared with Fig.~\ref{fig2}.}
  \label{fig3}
  \vspace*{0.3cm}
\end{figure*}

The baroclinic wave solutions to equation~(\ref{baroclinic_Dimfreq})
are presented in Fig.~\ref{fig2} for a planet with HD209458b
parameters given in Table~1.  The left panel shows the growth rate,
$k\!\cdot\!\Im\mathfrak{m}\{c\}$, as a function of wavelength
$2\pi\,k^{-1}$ at several different latitudes: $\phi = (60\degr,
45\degr, 35\degr, 25\degr)$.  The growth rates are labelled `HD60'
(red), `HD45' (green), `HD35' (yellow) and `HD25' (blue),
respectively.  The right panel shows the corresponding phase wave
speeds $\Re\mathfrak{e}\{c\}$.  Note here that unstable baroclinic
waves travel westward relative to the mean flow ($\Re\mathfrak{e}\{c\}
< 0$).

From Fig.~\ref{fig2} it can be seen that the wavelength of the most
unstable mode at $\phi = 60\degr$ is $1.7\, \times\, 10^8$~m,
corresponding to 1.8~undulations (i.e. $\sim$\,2 crests and troughs
each) at this latitude.  The growth rate of the instability is
$3.1\,\tau^{-1}$, where $\tau = 2\pi\,\Omega^{-1}$ is the planetary
rotation time.  At $\phi = 45\degr$ and $\phi = 35\degr$, the most
unstable modes correspond to 2.2 and 2.3~undulations around their
respective latitude circles and with growth rates $2.3\,\tau^{-1}$ and
$1.5\,\tau^{-1}$, respectively.  Hence, jets centred at lower
latitudes have increased growth times with modestly increased
wavelengths of the most unstable mode.  Significantly, linear analysis
predicts stability for jets located at or equator-ward of $\phi =
28\degr$ (see e.g. the flat, blue curve labelled `HD25').

To illustrate the dependence of the growth rate and phase speed on the
characteristic flow speed (or, equivalently, shear strength), we also
present in Fig.~\ref{fig2} result obtained for the case with $U_0 =
200$~m~s$^{-1}$ at $\phi = 45\degr$ (black curve labelled `HD45L').
Comparing the `HD45' (green) and `HD45L' (black) curves, we see
immediately that the growth rate of the most unstable mode decreases
significantly for the smaller $U_0$ case.  The instability takes
$\sim$\,4 times longer to develop in the weaker speed/shear case.  We
also note that the wavelength of the most unstable mode decreases
slightly.  Hence, as $U_0$ decreases, the number of undulations
increases for a jet located at a given latitude.

The qualitative behaviour described above is not restricted to
HD209458b.  It applies to any planet that has a meridional temperature
gradient.  To illustrate the general applicability of our results, we
present in Fig.~\ref{fig3} the growth rates and phase speeds at $\phi
= 45\degr$ for several planets: Earth, Jupiter and GJ436b (red, green
and black curves, respectively).  For the Earth, the wavelength of the
most unstable mode is 4100~km, corresponding to $\sim\! 7$ undulations
at midlatitude, with growth rate of $1.6\,\tau^{-1}$ (i.e. growth time
of 15~hours).  This is consistent with many studies of baroclinic
instability on the Earth \citep[e.g.][]{Thorncroft93,Polvani04}.  The
corresponding values of undulations for Jupiter and GJ436b are
$\sim$\,43 and $\sim$\,1 with growth rates $0.48\,\tau^{-1}$ and
$0.56\,\tau^{-1}$, respectively.  Accordingly, if baroclinic
instability occurs on these planets, it appears Jupiter simulations
must be of very high resolution to capture the instability.  On the
other hand, the instability at the midlatitude of GJ436b would clearly
be of planetary scale and thus may lead to a possible observable
variability signal for this planet on a timescale of
$\sim\,$1.8~planetary rotations.  Note that the phase speeds of the
unstable baroclinic waves on Earth and Jupiter are very small (close
to zero) compared to those on the extrasolar planets, HD209458b and
GJ436b.

We have also carried out linear growth rate analysis with the
two-layer QG model for HD209458b and have compared the results with
those from the primitive equations model, presented above.  In the two
models, the growth rates at high latitudes and midlatitudes are
equivalent to within $5$ per cent.  However, at low latitudes, the QG
model overestimates the growth rates by approximately 25 per cent.
Moreover, the QG model predicts instability down to $\phi = 23\degr$,
whereas the primitive equations model predicts instability only down
to $\phi = 28\degr$.  Below these latitudes, both models predict
stability.  Thus, ageostrophy appears to provide a stabilizing factor
in this case.  Given that inertia-gravity waves are not filtered in
the primitive equations model (as they are in the QG model), the
enhanced stability may be due to the gravity waves `leaking away' some
of the energy that drive the instability.

\subsection{Limitations} 

The preceding analysis is highly idealised.  Therefore, it has
limitations.  For example, in general, planetary jets possess a
three-dimensional structure -- with concurrent vertical and meridional
shears, as well as zonal asymmetry.  Also, the atmosphere is
continuously stratified.  One effect of a two-layer discretisation
with uniform zonal flow in each layer is the inability to capture the
symmetry breaking between eastward and westward jets.  These
limitations are discussed more in detail below.

In flows with both vertical and horizontal shears, the growth of
unstable baroclinic waves may be suppressed by the `barotropic
governor' effect \citep[e.g.][]{James87, Nakamura93I, Pedlosky64}.
The effect is not fundamentally related to the sign of the jet, but a
key ingredient is a counter-gradient eddy momentum flux
$\overline{u'v'}$ generated under a horizontally sheared flow; here
the overbar indicates a zonal average.  The shear and the momentum
flux reinforce each other to distort the meridional structure of the
wave, suppressing the growth rate and shortening the wavelength of the
most unstable mode.  Thus, the full non-linear evolution of the
instability exhibits lower growth rates and shorter wavelengths,
compared with those indicated by the linear analysis presented in this
section (see section~\ref{non-linear}).

The atmosphere is also continuously stratified.  A representation more
realistic than a two-layer model changes the instability properties
described in this section.  The main change is that the short-wave and
long-wave cut-offs in the two-layer representation (see
Figs.~\ref{fig2} and \ref{fig3}) no longer exist in the continuum of
unstable modes \citep[e.g.][]{Charney47,Green60,Kuo79}.  The retained
modes (Charney and Green modes, discussed below) are not expected to
change qualitatively the asymptotic behaviour of the instability.
However, they do provide additional modes for wave-wave interaction
during the non-linear growth phase -- hence, affect the details of the
evolution; this may be significant for finite-time variability.

Another limitation of the linear model presented is that it does not
distinguish between the signs of the jet (or shear).  This is because
symmetry is preserved under the interchange of the shear sign, given
the laterally uniform flow; hence, distinction between the two signs
is not expected.  This is in contrast to the flow used in the
non-linear calculation (section~\ref{non-linear}), in which the growth
rate for an unstable westward (negative shear) jet is smaller than
that for the unstable eastward (positive shear) jet at the same
latitude.  The two signed flows behave differently in this case
because of the change in the sign of the jet curvature.  Furthermore,
a westward jet has only one unstable mode (Charney mode) as opposed to
an eastward jet, which has an infinite number of unstable modes (Green
modes).

A similar observation has been made by \citet{Wang89}, who observed a
difference between eastward- and westward-sheared baroclinic flows in
the Charney model (continuously stratified QG model on the
$\beta$-plane).  He has pointed out that the maximum growth rate for
the absolute value of non-dimensional shear is substantially smaller
for a flow with westward shear than a flow with eastward shear.
Moreover, while the eastward jet is baroclinically unstable for any
value of vertical shear $\Lambda$, the westward jet is unstable only
if
\begin{equation}\label{crit_shear}
  \Lambda\ < \ -\frac{\beta \, N^2 \, H}{f_0^2}\, .
\end{equation}
Note that, at $\phi = 45\degr$, the critical shear for HD209458b
parameters used in this work is: $\Lambda_c = -7.9 \times
10^{-4}$~s$^{-1}$.  The shear of the unstable westward jet described
in section~\ref{westward} is $\Lambda = -1.7 \times 10^{-3}$~s$^{-1}$,
consistent with~(\ref{crit_shear}).

\section{Non-Linear Evolution}\label{non-linear}

\subsection{Numerical Model}

To study the full non-linear evolution, we use a well-tested parallel
pseudospectral model, BOB\footnote{`Built On Beowolf'}
\citep{Scott03}.  This model solves equations~(\ref{PE}) in spherical
geometry, subject to the boundary conditions~(\ref{BC}).  As in many
models, the equations in the vorticity--divergence form are solved,
where (relative) vorticity is $\zeta({\bf x},t) = {\bf
  k}\cdot\nabla\times{\bf v}$ and divergence is $\delta({\bf x},t) =
\nabla\cdot{\bf v}$.  The equations in this form are more amenable for
the spectral transform method \citep{Orszag70,Eliasen70,Canuto88}.
Domain decomposed spectral transform algorithm is used in the
horizontal direction and standard second order finite difference
scheme is used in the vertical direction.  The latter direction is in
pressure coordinates in the numerical model.

To follow the evolution over long duration, explicit dissipation is
applied so that artificial accumulation of energy at small scales is
prevented \citep[see e.g. ][]{Cho96}.  The dissipation is in the form
of a linear superviscosity operator, $-\nu\nabla^4(\cdot)$, applied to
the prognostic variables, $\{\zeta, \delta, \theta\}$; here $\nu$ is
constant.  A small Robert--Asselin time filter $\epsilon$
\citep{Robert66, Asselin} is applied, at every time step and in each
layer, to filter the computational mode arising from using a
second-order time-marching scheme \citep[see e.g.][]{Thrastarson11}.
No other numerical dissipators, fixers, stabilisers or filters are
used in performing the simulations.

\subsection{Model Setup}\label{setup}
 
To study the non-linear evolution of baroclinic instability, we
initialize our model with an idealized jet that satisfies the
necessary condition for baroclinic instability, the
Charney-Stern-Pedlosky condition described in section~\ref{linear}.
The jet is initially set to be either eastward or westward, and
centred at a latitude between $0\degr$ to $60\degr\,$N.  A large
number of simulations have been performed for this study, carefully
varying each parameter (jet location, strength, shear, profile,
direction as well as domain size, etc.) in an independent series of
simulations.  A very small subset of these runs, which we use for
discussions in sections~\ref{setup} to \ref{equatorial}, is given in
Table~\ref{table-runs}.  The set illustrates the basic points we wish
to make.

\begin{figure*}
  \vspace*{.3cm}
  \includegraphics[scale=0.9]{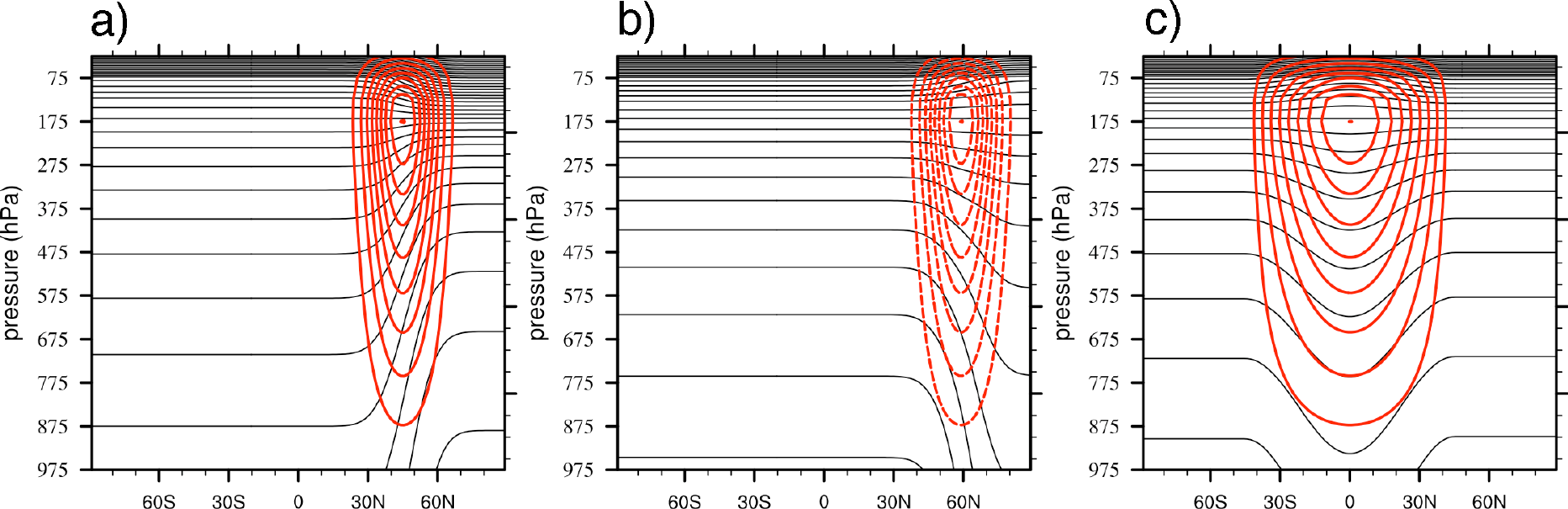}
  \vspace*{.3cm}
  \caption{The basic state zonal wind $u_0$ [m~s$^{-1}$] (red) and potential
    temperature $\theta_0$ [K] (black) as a function of latitude and pressure
    for runs (see Table~\ref{table-runs}): a)~E45N, b)~W60N, and 
    c)~EEQ.  Contour interval for the zonal wind is 100 m~s$^{-1}$ and
    for the potential temperature 100~K.  Negative contours are
    dashed.}
  \vspace*{.2cm}
  \label{fig4}
\end{figure*}

All the jets are initially non-linearly balanced so that a
self-consistent background temperature structure is generated
(Fig.~\ref{fig4}).  The jets are then bumped at the beginning of the
simulation by an infinitesimal temperature disturbance which is
independent of altitude, a barotropic `heat bump', and allowed to
evolve freely thereafter.  The setup is chosen to be similar to that
in \citet{Polvani04} for validation and comparison purposes.  For
example, following that work, the initial zonal flow $u_0$ in our runs
is, in general,
\begin{equation}\label{jet_profile}
  u_0(\phi,p)\ =\
  \begin{cases}
    U\,\sin^m[\pi \sin^2\left(\phi-\phi_0\right)]\,F(z^*), &
    \hspace*{.05cm} \phi_0<\phi < \phi_{\text{T}}\\
    0\, , & \hspace*{.3cm} \text{otherwise}\, .
  \end{cases}
\end{equation} 
Here 
\begin{equation}
  F(z^*)\ =\ \frac{1}{2}\left[1 - \tanh^3\left(\frac{z^* - z_0}{\Delta
        z_0}\right)\right]\sin\left(\frac{\pi z^*}{z_{1}}\right)
\end{equation}
with $z^* = -H \log (p/p_{\rm r})$, and $\phi_0$ and $\phi_{\text{T}}$
are taken to be the following: $\phi_0 = 0$ and $\phi_{\text{T}}=
\pi/2$ for jets centred at midlatitude (E45N and E45N2b), $\phi_0=
\pi/12$ and $\phi_{\text{T}} = \pi/2$ for jets centred at $60\degr$N
(W60N) and $\phi_0 = -\pi/4$ and $\phi_{\text{T}} = \pi/4$ for jets
centred on the equator (EEQ).  The typical values of the parameters
are: $U = \pm 1000$~m~s$^{-1}$, $z_{0} = 1823$~km, $z_{1} = 2486$~km,
$\Delta z_{0} = 414$~km, $H = 580$~km, and $p_{\rm r} = 10^5$~Pa (=
1~bar).  The latitudinal width of the jet is determined by $m$ in
(\ref{jet_profile}), where $m = 3$ corresponds to a jet width of
$\sim\,$40$\degr$ (Fig.~\ref{fig4}a and Fig.~\ref{fig4}b) and $m =
1/2$ to a width of $\sim\,$85$\degr$ (Fig.~\ref{fig4}c).  To discuss
jets that closely match those produced in current GCM simulations of
extrasolar giant planets, we present runs which are initialized with
wider ($m = 1/2$) jets in the equatorial region and narrower jets ($m
= 3$) poleward of $45\degr\,$N.

The basic state temperature, $T_0 = T_0(\phi,p)$, is obtained by
combining meridional momentum and hydrostatic balance equations:
\begin{equation}\label{gradient wind}
  \frac{\partial T_0}{\partial \phi}\ =\ 
  -\frac{H}{\cal R}\,(R_pf + 2u_0 \tan
  \phi)\,\frac{\partial u_0\ }{\partial z^*}\, .
\end{equation}
Integrating (\ref{gradient wind}) results in a temperature
distribution that is in non-linear, gradient wind balance with the
specified jet.  Here we have used a reference temperature of 1500~K as
the constant of integration.  The value is consistent with initial
conditions and results of many GCM calculations.  The basic state flow
$u_0(\phi,p)$ and potential temperature $\theta_0(\phi,p)$ for runs
E45N (eastward midlatitude jet), W60N (westward high latitude jet) and
EEQ (wide eastward equatorial jet) are shown in Fig.~\ref{fig4}.
Recall that $\theta_0$ is related to $T_0$ by $\theta_0 = T_0(p_{\rm
  r}/p)^\kappa$.  To catalyse the instability, $T_0$ is given a small
perturbation $T^\prime$ in the form of a localized bump at all
pressure levels such that
\begin{equation}
  T^\prime(\lambda,\phi)\ =\ 
  {\cal A}\,{\rm sech}^{2}\left[3 \, (\lambda - \lambda_0)\right]\,
  {\rm sech}^{2}\left[6(\phi-\phi_{0})\right]\, ,
\end{equation}
for $-\pi < \lambda < \pi$.  Here ${\cal A} = 1$~K and $(\phi_{0},
\lambda_0)$ represents the jet centre (latitude,\,longitude).
 
Our vertical domain, which typically extends from 1 to 10$^{-3}$~bar,
is resolved by 20 equally spaced pressure levels.  The horizontal
resolution of results presented in section~\ref{base case} to
section~\ref{equatorial} is T170, or 170 sectoral modes and 170 total
modes in the spectral expansion \citep[see e.g. ][]{Thrastarson11}.
The resolution is designated `T170L20'.  The inverse transformation is
performed on to a 512$\times$256 Gaussian grid covering the entire
globe.  The grid size is chosen for de-aliasing \citep[][and
references therein]{Canuto88}.  Equations~(\ref{PE}) are integrated
for up to 60\,$\tau$ (i.e. 60 planetary rotations) with $\nu =
6\times10^{19}$~m$^4$~s$^{-1}$.  A timestep size, $\Delta t = 30$~s,
and a Robert-Asselin coefficient, $\epsilon = 0.01$, are used for the
time integration.

As already mentioned, the choice of our initial conditions is partly
motivated by current GCM results of hot extrasolar giant planet
atmospheres.  These studies suggest typical flow speeds of
$O(100-3000$~m~s$^{-1})$ and zonal flow consisting of up to $\sim\,$3
jets -- often a broad equatorial eastward jet and a smaller amplitude
narrower westward jet at a higher latitude on both northern and
southern hemispheres
\citep[e.g.][]{Showman08,Rauscher10,Thrastarson10,Heng11}. The
altitudinal and latitudinal profiles used here roughly mimic those
presented in fig.~9 of \citet{Showman08} and fig.~3 of
\citet{Rauscher10}.

In what follows, we first describe the evolution of the midlatitude
{\it eastward} jet (run E45N).  Although such a jet is not commonly
observed in current simulations of hot extrasolar giant planets,
reviewing this case is useful because it allows the present work to be
compared with analogous studies -- and observations -- of the Earth
and because it allows a baseline to be constructed for other initial
conditions presented here, namely the high-latitude westward and
equatorial eastward jets that match more closely with aforementioned
extrasolar planet simulations.

\begin{table}
  \centering
  \caption{Summary of jet configurations discussed: $m$ is a 
    parameter that controls the jet width [see 
    equation~(\ref{jet_profile})].  Note, in run E45N2b the 
    bottom boundary is set at $p = 2$~bar and the vertical 
    structure function $F(z^*)$ in equation~(\ref{jet_profile})
    is specified as $ F(z^{**}) = 
    \{\,1 - \tanh^8[\,(z^{**} - z_{\rm 2b})\,/\,\Delta z_0\,]\,\}\,
    \sin^4(\pi\,z^{**}/z_1),$
    where $z^{**} = -H \log [(p+p_0)/p_{\rm r}]$, 
    with $p_0 = 60$~hPa and $z_{\rm 2b} = 900$~km.}
    \label{table-runs}
    \begin{tabular}{lccl} \hline \hline
      \\
      Run   & Width ($m$) &  Latitude & Direction \\
      \\
      \hline
      \\
      E45N   & 3.0  & $45\degr\,$N   & East \\
      E45N2b  & 3.0  & $45\degr\,$N   & East \\
      W60N   & 3.0  & $60\degr\,$N   & West \\
      EEQ    & 0.5  & $0\degr$       & East \\
    \\
    \hline\hline \\
  \end{tabular}
\end{table}

\subsection{Paradigm Case}\label{base case}

\begin{figure}
  \vspace*{.3cm}\hspace*{.1cm}
  \includegraphics[scale=0.42]{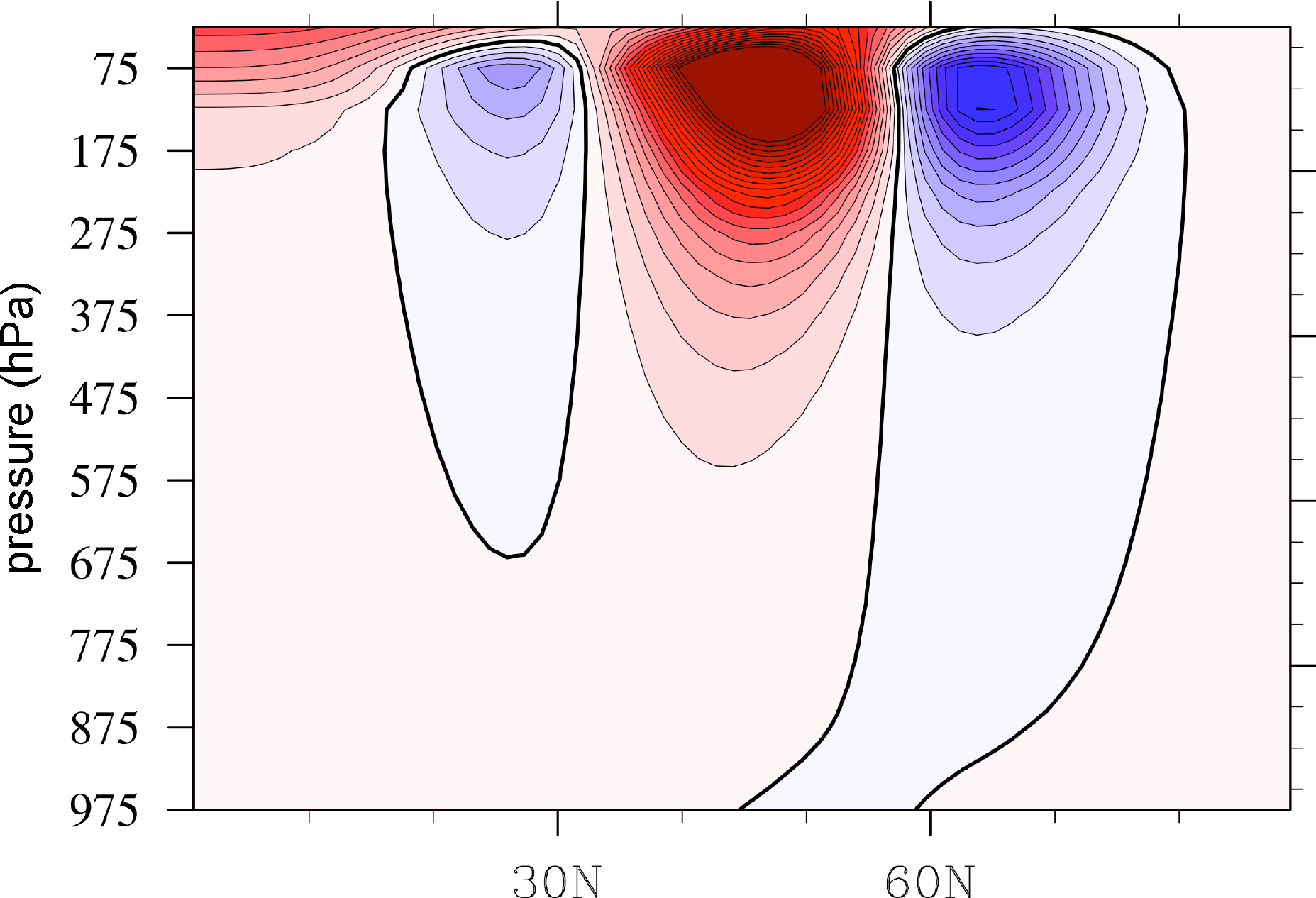}
  \vspace*{.3cm}
  \caption{Meridional cross-section of the meridional potential
    vorticity gradient $(\partial q_0 / \partial y)_{\theta}$ for run E45N
    (northern hemisphere).  Maximum and minimum values are $\pm\, 3
    \times 10^{-12}$~K~m~kg$^{-1}$~s$^{-1}$ with contour interval $2
    \times 10^{-13}$~K~m~kg$^{-1}$~s$^{-1}$.  Negative values are in
    blue and positive are in red. The zero contour is drawn with
    double thickness.}
  \label{fig5}
  \vspace*{.2cm}
\end{figure}

\begin{figure*}
  \vspace*{.3cm}
  $\begin{array}{cc}
    \includegraphics[height=21cm]{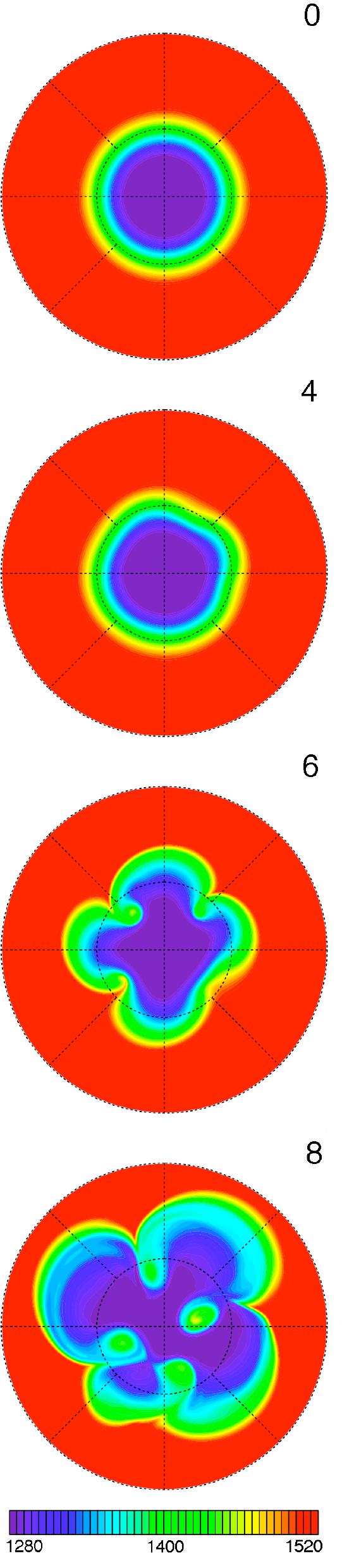}
    \hspace*{1.5cm}
    \includegraphics[height=21cm]{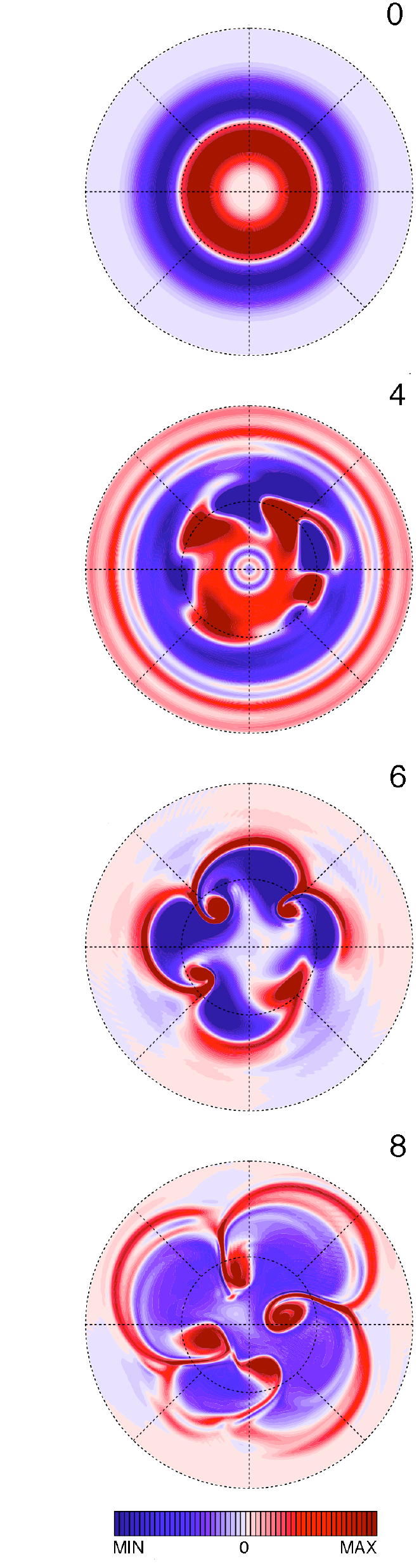}
  \end{array}$\\*[.3cm]
  \caption{Temperature $T$ (left) and relative vorticity $\zeta$
    (right) from run E45N in polar stereographic view, centred on the
    north pole.  The fields are shown at the 975~hPa pressure level
    for $\tau = 0$ to $\tau = 8$.  Maximum and minimum values for $T$
    are 1280~K and 1520~K, respectively, with contour interval 6~K.
    For $\zeta$, the maximum and minimum values are $\pm 5 \times
    10^{-7}$~s$^{-1}$ ($\tau = 0$), $\pm 1 \times 10^{-6}$~s$^{-1}$
    ($\tau = 4$), $\pm 1 \times 10^{-5}$~s$^{-1}$ ($\tau = 6$) and
    $\pm 4 \times 10^{-5}$~s$^{-1}$ ($\tau = 8$); the contour
    intervals are, respectively, $2 \times 10^{-8}$~s$^{-1}$, $4
    \times 10^{-8}$~s$^{-1}$, $4 \times 10^{-7}$~s$^{-1}$ and $1.6
    \times 10^{-6}$~s$^{-1}$.  The spectral resolution of this
    simulation is T170L20 (see text).  Note the large, nearly two
    orders of magnitude, change in the magnitude of $\zeta$ during the
    evolution -- as well as the formation of sharp fronts and coherent
    vortices, particularly at $\tau = 6$ and $\tau = 8$.}
  \label{fig6}
\end{figure*}

Run E45N is the `paradigm case'.  It illustrates a typical non-linear
evolution of a perturbed, marginally stable\footnote{While the jets
  satisfy the (necessary) condition for instability, they require an
  initial perturbation to evolve: they are perfectly stable without
  the perturbation.}  baroclinic jet on a hot extrasolar giant planet
in numerical simulations with high resolution.  The jet is zonally
symmetric and eastward with speed 1000~m~s$^{-1}$ at the jet core and
decaying to zero at the periphery (see Fig.~\ref{fig4}a).  It meets
the necessary conditions (i), (ii) and (iv) for baroclinic
instability, defined in section~\ref{linear}, with (i) and (ii) only
weakly satisfied.  This can be seen from Fig.~\ref{fig5}, which shows
$(\partial q_0 / \partial y)_{\theta}$ evaluated on an isentrope as a
function of $\phi$ and $p$. Note, here $q_0$ is the potential
vorticity defined on isobars,
    \[q_0(\phi,p)\ =\ -g (f \bf{k} +
    \nabla\!\times\!\bf{v_0})\!\cdot\!  \nabla\theta_0 \, \, ,
    \] where $\nabla$ is the three-dimensional gradient operator in
    $(\lambda,\phi,p)$ space; and, $(\partial q_0 / \partial
    y)_{\theta}$ is a derivative taken along an isentrope such that
\[\left(\frac{\partial q_0 }{ \partial y}\right)_{\theta}\ =\ 
\left(\frac{\partial q_0}{\partial y}\right)_p - \left(\frac{\partial
    \theta_0}{\partial y}\right)_p\!\left(\frac{\partial
    \theta_0}{\partial p}\right)^{-1}\!\frac{\partial q_0}{\partial
  p}\,\, ,
\]
where $y=R_p\phi$ and $[\partial (\,\cdot\,) /\partial y]_p$ is the
derivative taken on an isobar. Other cases, with jets of different
sign or location, are to be compared with this one.  Fig.~\ref{fig6}
presents the evolution of $T$ (left column) and $\zeta$ (right column)
fields at the $p = 975$~hPa surface from run E45N, for $\tau = 0$ to
$\tau = 8$.  The fields near the reference pressure level are shown
since the kinetic energy is the maximum at the lower boundary for jet
profiles shown in Fig.~\ref{fig4}, similar to \citet{Gall76} and
\citet{Simons72}.  Note that for these jets $T\approx\theta$ at this
pressure level.

In Fig.~\ref{fig6}, the perturbed jet undergoes initially a period of
linear growth ($\tau \ll 4$), when the most unstable mode emerges.  At
this early stage, the $T$ field shows a small-amplitude perturbation
from zonal symmetry.  The $\zeta$ field, on the other hand, is much
more dynamic.  At $\tau\! =\! 4$, finite-amplitude wave breaking in
the $\zeta$ field is already clearly evident, and the perturbation in
this field is characterized by a distinct northwest--southeast tilt on
the poleward side of the jet and southwest--northeast tilt on the
equatorward side of the jet.  The enhancement of the tilt proceeds
concomitantly with the barotropic component of the flow, which
generates negative meridional flux of the eddy zonal momentum (i.e.
$\overline{u'v'} < 0$) on the poleward flank of the jet and positive
meridional flux of the eddy zonal momentum (i.e.  $\overline{u'v'} >
0$) on the equatorward flank of the jet \citep[see
e.g.][]{Nakamura93II}.

\begin{figure}
  \hspace*{-.7cm}
  \includegraphics[scale=0.34]{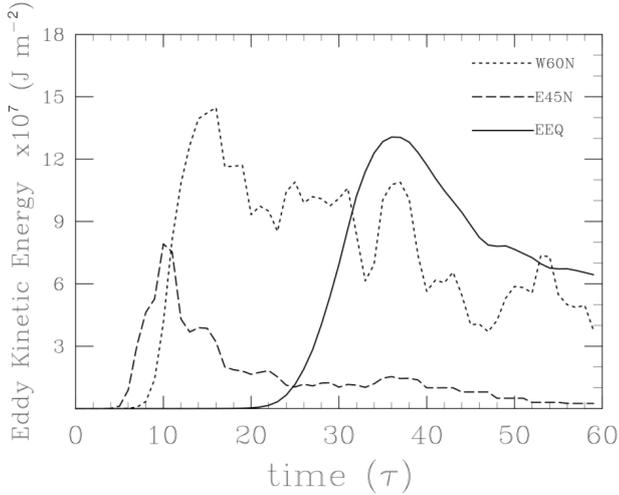}\\*[-.6cm]
  \caption{Evolution of globally-averaged eddy kinetic energy (per
    area) [J~m$^{-2}$] for runs E45N (dashed line), W60N (dotted
    line) and EEQ (solid line).  The eddy kinetic energy for EEQ has
    been multiplied by a factor of 50.}
  \label{fig7}
\end{figure}

During this early stage of the evolution, conversion of available
potential energy (\!\APE\!) into eddy kinetic energy (\!\EKE\!) slowly
begins, as can be seen from Fig.~\ref{fig7}.  These quantities are
defined:
\begin{eqnarray}\label{APE}
  \lefteqn{\mbox{\it APE}\  =\ 
    -  \int_S \int_0^{p_r}\!
    \frac{p^{\kappa -1} \!\ {\cal R}}{2g \!\ p_r^{\ \kappa}}\!\
    (\theta'')^2 \left(\frac{\partial \hat{\theta}}{\partial p}\right)^{-1} \!
    \! dp\, dA}\\
  \lefteqn{\mbox{\it EKE}\  =\ \ \
     \int_S \int_0^{p_r}\!
    \frac{1}{2g}\left[(u')^2+(v')^2\right] dp\, dA\, , }\label{EKE}
\end{eqnarray}
where $\theta''$ is the deviation of $\theta$ from its isobaric
average $\hat{\theta}$, and the integrations are over the surface area
$A$ and pressure $p$.  The baroclinic instability taps the \APE to
drive the eddy motions.

At $\tau\approx 5$ a rapid non-linear development ensues in both
fields.  Note, for example, the scale change in the $\zeta$ field at
$\tau =6$.  A large amplitude wave can now also be clearly seen in the
$T$ field.  In both fields, sharp frontal features form.  These
dynamically-significant sharp features require very high resolution to
capture faithfully.  This will be discussed more in detail in
section~\ref{convergence}.  By $\tau = 8$ sharp temperature gradients
trail out around the anticyclonic region (large `clover-leaf' shaped
area of negative vorticity, shaded in blue), forming curved baroclinic
fronts.  At this time, the \EKE is well into its non-linear growth
stage.  Note the pools of warm air that have been pinched off
(cyclonic vortices embedded in the anticyclonic region), intruding
into the high latitudes.  Simultaneously, broad regions of cool air
spread into the tropical region from higher latitude (i.e. negative
heat transport).  Thus, the original equator--pole temperature
gradient is significantly reduced by the instability.

\begin{figure}
  \vspace*{.5cm}
  \includegraphics[scale=0.71]{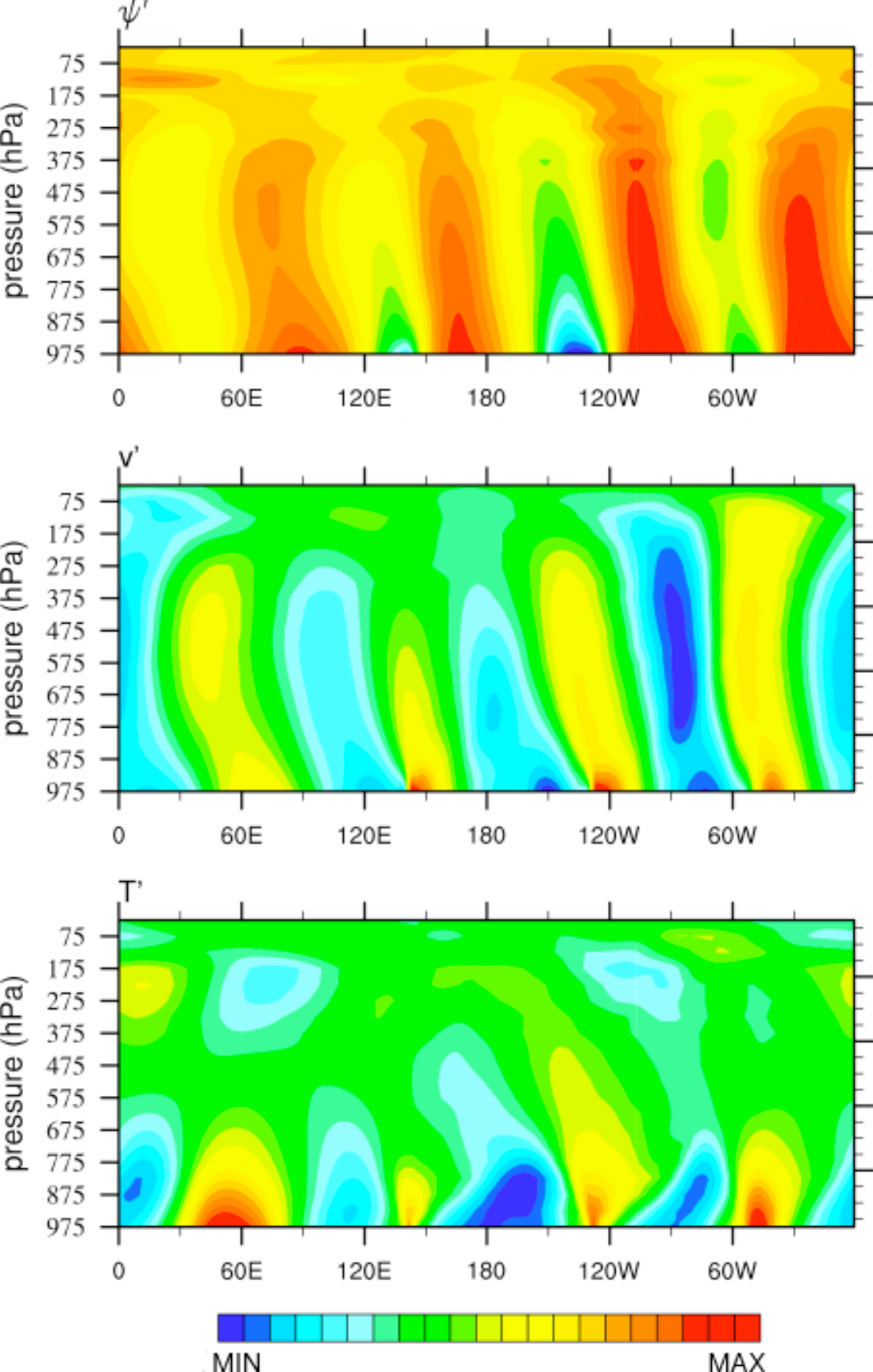}\\*[.1cm]
  \caption{Perturbation streamfunction $\psi^\prime$ (top),
    perturbation meridional velocity $v^\prime$ (middle) and
    perturbation temperature $T^\prime$ (bottom) at midlatitude as a
    function of pressure and longitude at $\tau = 6$ for run E45N.
    Contour intervals are: $-24 \times 10^8$~m$^2$~s$^{-1}$ to $11 \times
    10^8$~m$^2$~s$^{-1}$ in steps of $10^8$~m$^2$~s$^{-1}$,
    $-180$~m~s$^{-1}$ to 320~m~s$^{-1}$ in steps of 20~m~s$^{-1}$ and
    $-75$~K to 125~K in steps of 5~K, respectively.  Note,
    $\psi^\prime$ and $v^\prime$ tilt westward with height and
    $T^\prime$ tilts eastward with height, signifying meridional
    transport of heat and reduction of equator-pole temperature
    gradient.}
  \label{fig8}
\end{figure}

The poleward heat transport can be checked against linear theory for
eddy transport \citep[see e.g. ][]{Holton, Vallis} by examining the
zonal cross-sections of streamfunction, meridional velocity and
temperature perturbations:\, $\psi'$, $v'$ and $T'$, respectively.
The cross-sections at midlatitude are shown in Fig.~\ref{fig8}.  As
can be seen, $\psi'$ and $v'$ tilt westward with height and $T'$ tilts
eastward with height, demonstrating that heat transport is taking
place.  Note that, in the case of baroclinically unstable westward jet
at the same latitude, the directions of the tilts are reversed.  This
is because gradient wind balance produces a temperature distribution
that is warmer at the poles than at the equator, as can be seen in
Fig.~\ref{fig4}b.  This results in an equatorward transport of heat by
the eddies (section~\ref{westward}).

\begin{figure*}
\vspace*{.5cm}
$\begin{array}{cc}
    \hspace*{-.1cm}
    \includegraphics[height=11cm]{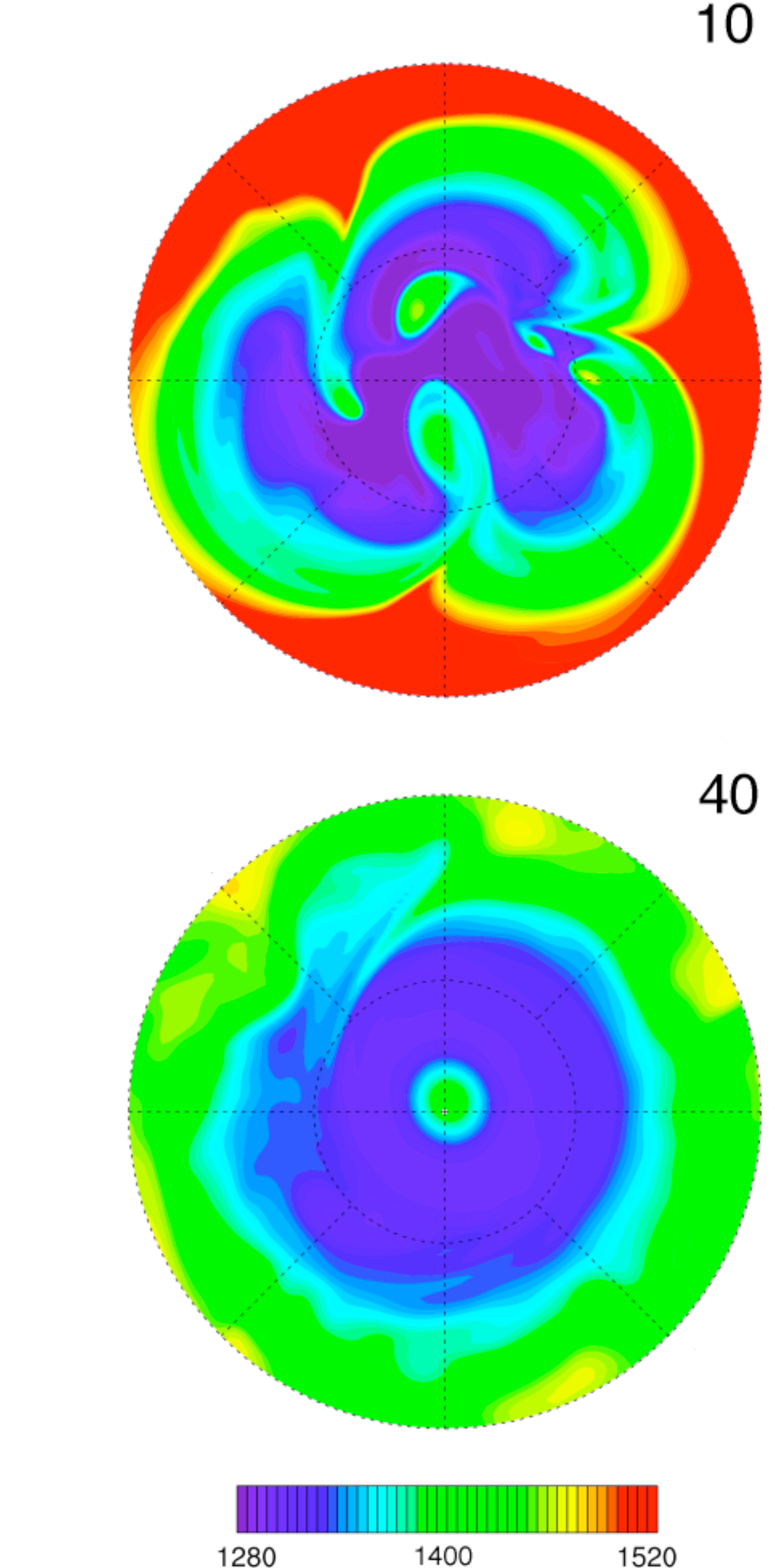}
    \hspace*{2.2cm}
    \includegraphics[height=11cm]{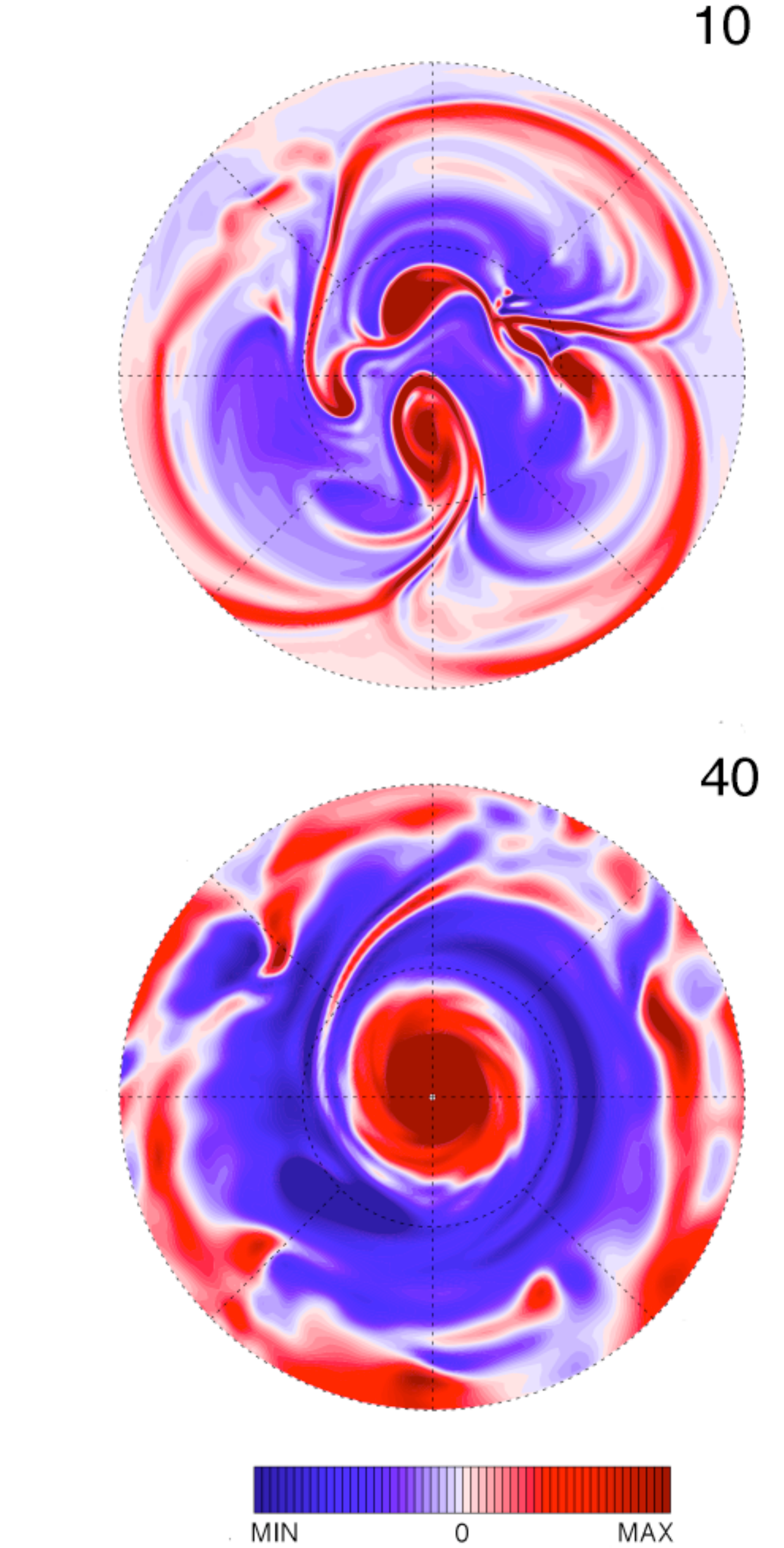}
  \end{array}$
\vspace*{.4cm}
\caption{Same as Fig.~\ref{fig6} but for $\tau = 10$ and $\tau =
  40$. Contour interval for temperature is 6~K.  The maximum and
  minimum contours for relative vorticity are $\pm 4 \times
  10^{-5}$~s$^{-1}$ at $\tau = 10$ and $\pm 10^{-5}$~s$^{-1}$ at $\tau =
  40$.  The respective contour intervals are $1.6 \times
  10^{-6}$~s$^{-1}$ and $4 \times 10^{-7}$~s$^{-1}$.}
  \label{fig9}
\end{figure*}

The long-time evolution of the run presented in Fig.~\ref{fig6} is
illustrated in Fig.~\ref{fig9} ($\tau = 10$ and $\tau = 40$).  By
$\tau \approx 40$ the $T$ and $\zeta$ fields have organised into
essentially zonal structures and eddy activity has mostly ceased.  The
cyclones that have emerged from the baroclinic wave breaking, strongly
interact ($\tau = 10$) and ultimately merge into an unsteady cyclonic
polar vortex ($\tau = 40$).  A similar `end-state', resulting from
vortex mergers, has been observed in HD209458b simulations of
\citet{Cho03}.

The long-range interaction of the like-signed vortices on hot
extrasolar planets is more pronounced than on the Earth (and other
cool, rapidly-rotating planets).  This can be explained by the much
larger Rossby deformation length scale, $\LD/R_p = {\cal O}(1)$, on
the hot extrasolar giant planet.  Larger $\LD$ means more robust
mergers and a more dynamic final vortex \citep{Cho03,Cho08a, Cho96}.
\citet{Scott11} has recently quantified this behaviour: merger and
poleward migration of cyclones ensues if the potential vorticity
anomaly $q'$ associated with a vortex exceeds the magnitude of the
planetary vorticity $2\Omega $ by $\sim\,$12 per cent.  In our case,
we find $(q'/2\Omega) \geq 1.19$ -- i.e. anomaly excess of 19 per cent
-- by $\tau = 6$, consistent with Scott's finding.

The temporal evolution of the global average \EKE (the dashed line for
run E45N and solid line for run EEQ in Fig.~\ref{fig7}) is typically
described as a `baroclinic growth\ --\ barotropic decay' cycle
\citep[e.g.][]{Sim79,Thorncroft93}.  During the cycle, conversion of
\APE to \EKE is impeded by a positive feedback between the horizontal
shear in the flow and the eddy momentum flux.  At $\tau \geq 10$, the
disturbances in run E45N are sheared out and \EKE is lost to the mean
flow through the Reynolds stresses more than it is gained through
baroclinic conversion.  The feedback is the main component in the
previously mentioned non-linear `barotropic governor effect', affected
by the horizontal shear in the jet, spherical geometry and ageostrophy
\citep[see e.g.][]{Nakamura93II}.

\begin{figure*}
  \vspace*{1.5cm}\hspace*{.1cm}
  \includegraphics[scale=0.9]{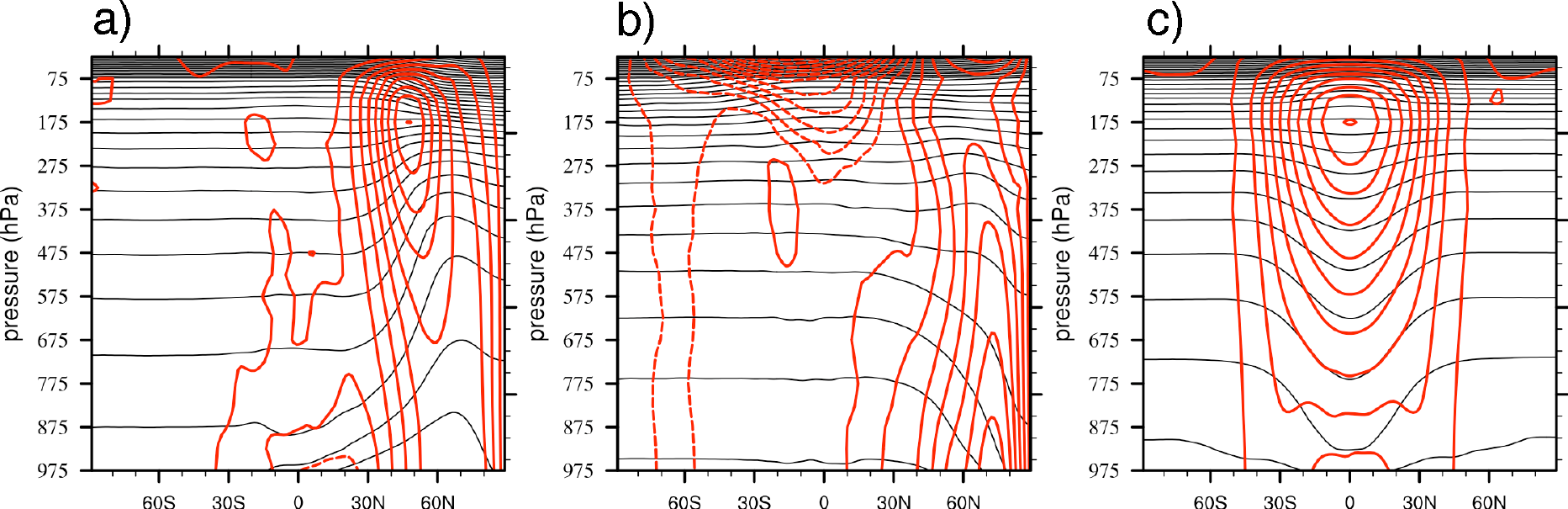}
  \vspace*{.3cm}
  \caption{Zonal mean zonal wind $\bar{u}$ (red) and potential temperature
    $\bar{\theta}$ (black) contours for runs a) E45N at $\tau = 30$, b) W60N
    at $\tau = 60$ and c) EEQ at $\tau = 60$.  Wind contour interval
    for runs E45N and EEQ is 100~m~s$^{-1}$ and for run W60N is
    50~m~s$^{-1}$.  Temperature contour interval is 100~K.  Negative
    (westward) wind contours are dashed. }
  \label{fig10}
\end{figure*}

The zonal mean zonal wind $\bar{u}$ and zonal mean zonal potential
temperature $\bar{\theta}$ at the end of the life-cycle is presented
in Fig.~\ref{fig10} (see panel {\it a} on the left for run E45N).  The
jet itself has becomes broader and more barotropic -- much like the
`LC1 life cycle' reported in \citet{Thorncroft93}.  Furthermore, the
meridional entropy gradient $\d\bar{\theta}/\d\phi$ is
significantly reduced compared to the initial state, particularly at
the lower levels in the domain (cf. Fig.~\ref{fig4}a). Much of the
\APE is taken up by the kinetic energy of the zonal mean flow and the
flow is accelerated there.  To quantify the accelerations, consider
the transformed Eulerian-mean zonal momentum equation
\citep[e.g.][]{AndrewsMac78}:
\begin{eqnarray*}\label{TEM}
  \frac{\partial \, \bar{u}}{\partial \, t}\  =\ -\left[
    \frac{1}{R_p\, \cos\phi}\frac{\partial}{\partial\, \phi}
    (\bar{u} \, \cos\phi) - f\right]\, \bar{v}^\ast\, - \,
  \frac{\partial \bar{u}}{\partial\,
    p}\,\bar{\omega}^\ast \ \\
  +\, \frac{1}{R_p \, \cos\phi}\, \nabla \cdot \bold{F}\, .
\end{eqnarray*}
Here $\bar{v}^\ast$ and $\bar{\omega}^\ast$ represent the
`residual' mean meridional circulation,
\begin{eqnarray*}\label{TEM_2}
  \bar{v}^\ast & \equiv & \bar{v} - \frac{\partial}{\partial p} 
  \left(\frac{\overline{v'\theta'}}
    {\partial \bar{\theta}/\partial p}\right) \\
  \bar{\omega}^\ast & \equiv & \bar{\omega} +
  \frac{1}{R_p \cos\phi}\frac{\partial}{\partial \phi}
  \left(\frac{\overline{v'\theta'}}{\partial \bar{\theta}/\partial
      p} \cos\phi \right)\, ,
\end{eqnarray*}
and ${\bf F}\! =\! (F_\phi, F_p)$ is the Eliassen-Palm (E-P) flux
vector with
\begin{eqnarray*}\label{TEM_3} 
  \lefteqn{F_\phi\  =\  R_p \cos\phi \left[-\overline{u'v'} +
      \left(\frac{\overline{v'\theta'}}{\overline{\partial \theta/\partial
            p}}\right)\!\!
      \left(\,\overline{\frac{\partial \bar{u}}{\partial p}}\,\right)\right]} \\
    \lefteqn{F_p\  =\   R_p \cos\phi 
      \left[(\zeta + f)\! \left(\frac{\overline{v'\theta'}}{\overline{\partial
            \theta / \partial p}}\right) - \overline{u'\omega'}\right]\, .}
\end{eqnarray*}

\begin{figure}
  \hspace*{-.5cm}
  \includegraphics[scale=0.33]{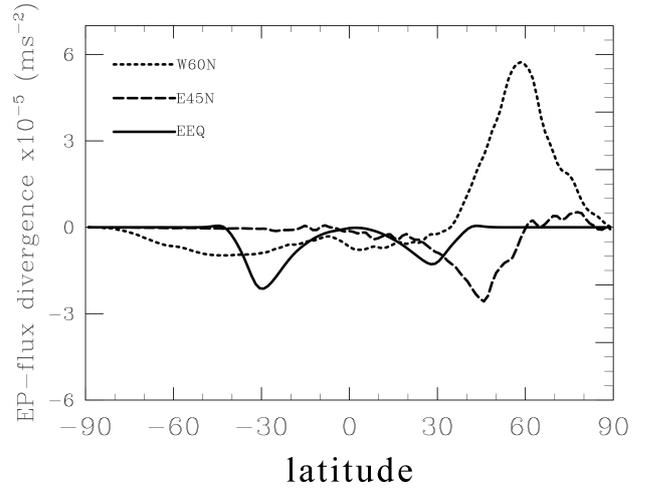}\\*[-.6cm]
  \caption{Vertically and temporally averaged divergence of
    Eliassen-Palm (E-P) flux [m~s$^{-2}$] for runs E45N (dashed line),
    W60N (dotted line) and EEQ (solid line) during the life-cycle in
    each run.  The EEQ curve has been multiplied by a factor of 100.}
  \label{fig11}
\end{figure}

The influence of eddies on the mean flow is measured by the E-P
fluxes: a convergent flux ($\nabla\! \cdot\! {\bf F} < 0$) corresponds
to the deceleration of the eastward flow and a divergent flux
($\nabla\!  \cdot\! {\bf F} > 0$) corresponds to acceleration.
Fig.~\ref{fig11} depicts vertically and temporally averaged E-P flux
divergence for E45N (dashed line) over the life-cycle.  The E-P fluxes
are divergent in the net on the poleward flank of the jet, where the
flow is accelerated, and (more strongly) convergent in the net on the
equatorial flank, where overall the flow speed is reduced from the
initial value (cf. Fig.~\ref{fig10}a ).

Finally, we note that the most unstable mode for this calculation is
$\sim\,$4 (see Fig.~\ref{fig6}).  As discussed earlier, the linear
theory of section~\ref{analysis} underestimates this number to
$\sim\,$2.  However, the full numerical simulation shows that the
simple linear theory is successful, at least qualitatively, in
capturing the behavior of the instability in the following sense: the
most unstable mode and the growth time for the baroclinic wave
amplitude for HD209458b are smaller than the corresponding quantities
for the Earth \citep[cf., for example,][]{Polvani04}.

\subsubsection{Lower Boundary}

As is well-known, boundary conditions are crucial in solving
differential equations.  Differences in the conditions, even in
relatively simple physical situations, can alter the admitted
solutions.  For example, new or modified modes can be introduced or
existing modes can be filtered by employing rigid boundary condition
(i.e. $w = 0$).  The lower boundary of the simulations discussed in
this paper is rigid and located at 1~bar for the most part.  In this
case, the vertical wind shear in the basic flow used is small, but
non-zero, at the bottom boundary and baroclinically unstable modes can
arise due to the presence of the boundary -- via condition (ii) of the
Charney-Stern-Pedlosky criteria.  However, the stably-stratified
radiative zone in hot extrasolar planet atmospheres may extend down to
perhaps as deep as 1000~bars \citep{Guillot02}.  Hence, the effects of
lowering the bottom boundary and preventing flow shear there require
careful consideration.

\begin{figure*}
  \vspace*{.5cm}
  $\begin{array}{cc}
    \includegraphics[height=7.2cm]{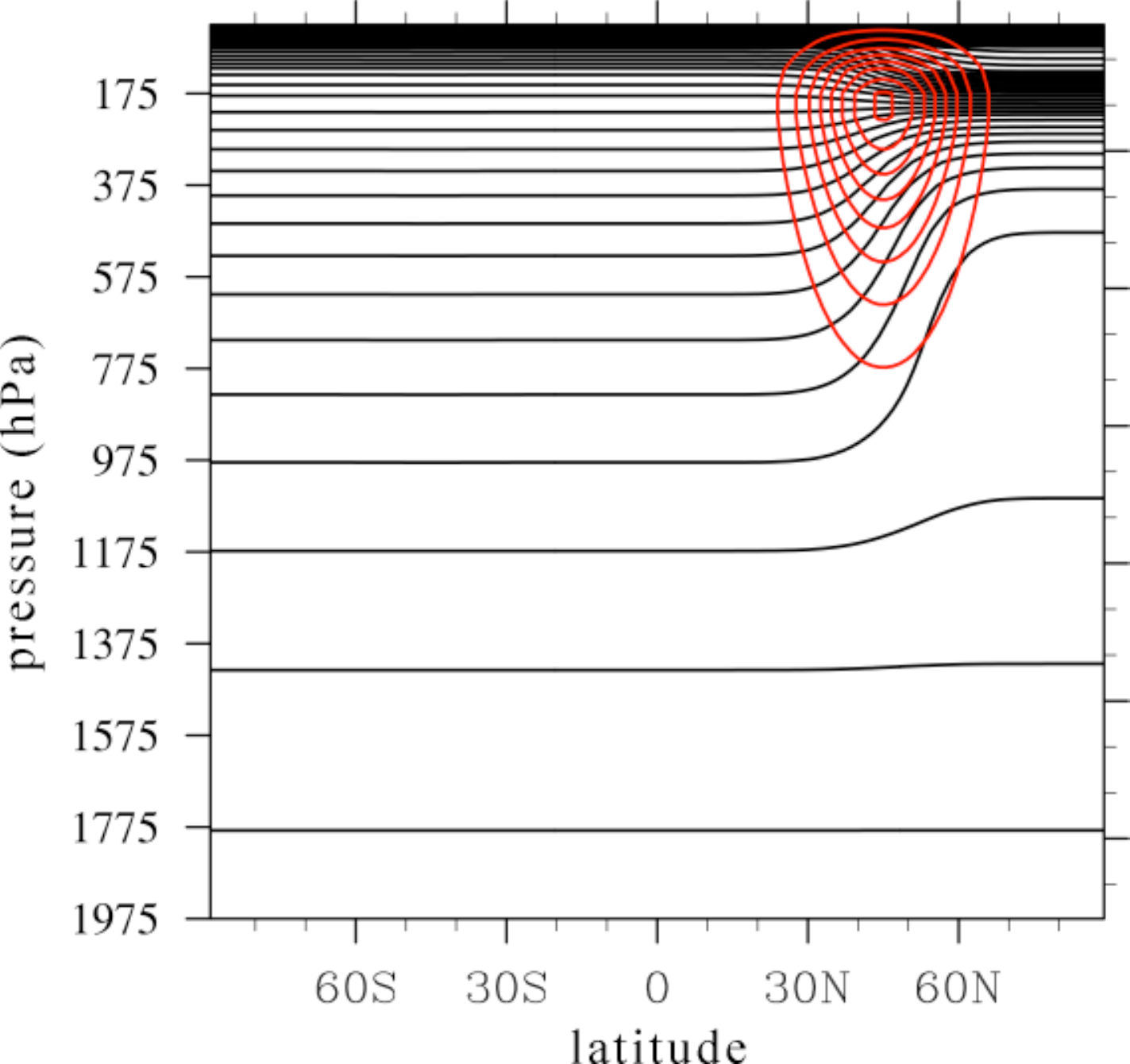}
    \hspace*{1.5cm}
    \includegraphics[height=7.2cm]{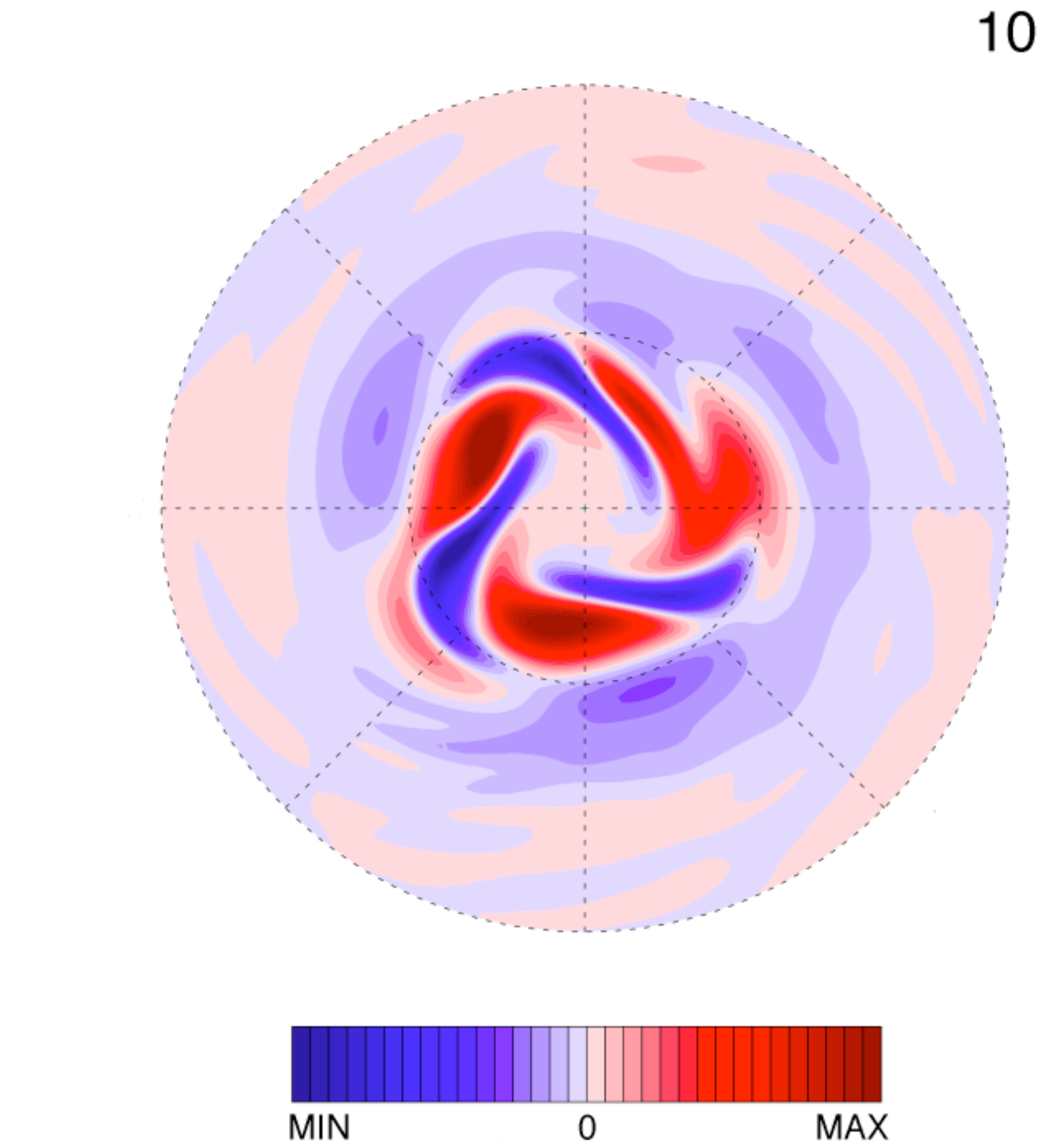}
  \end{array}$
  \vspace*{.3cm}
  \caption{Left: Basic state zonal wind $u_0$ (red) and potential
    temperature $\theta_0$ (black) for run E45N2b with the lower boundary
    extended to 2~bars.  Contour interval for the zonal wind is 100
    m~s$^{-1}$ and for the potential temperature 100~K.  Right:
    Relative vorticity $\zeta$ from run E45N2b at $\tau = 10$. The
    field is plotted at 975~hPa pressure level.  Maximum and minimum
    values are $\pm\, 6 \times 10^{-6}$~s$^{-1}$, with contour
    intervals of $4 \times 10^{-7}$~s$^{-1}$.}
  \label{fig12}
\end{figure*}

Fig.~\ref{fig12} presents a run (E45N2b) that is very similar to the
`paradigm case', but with the lower boundary of the calculation
extended down to 2~bars.  The jet is confined to pressures above the
1~bar level.  In doing so we remove the influence of the lower
boundary far enough away from the jet while still retaining an
adequate vertical resolution.  In the figure, $u_0$ and $\theta_0$ are
shown in the left panel.  Note, the jet profile shown in the figure
has a different vertical structure than the `paradigm case' jet.  This
is because balancing a jet with vertical structure given by
equation~(\ref{jet_profile}) to the isothermal reference state
produces a convectively unstable region in the computational domain,
causing the vertical coordinate to lose single-valuedness and the run
to immediately crash. The shown profile does not suffer from this.
Significantly, the Charney-Stern-Pedlosky criteria (iv) and (i) remain
satisfied for this profile.  It is crucial to understand here that,
once these criteria are met, it does not matter whether the lower
boundary is located at 10 or 1000~bars for the instability to occur.

The $\zeta$ field at $\tau\! =\! 10$ is shown in the right panel of
Fig.~\ref{fig12}. Although the evolution is now slightly altered from
the `paradigm case' (i.e. mode-3 is dominant, rather than mode-4), the
jet is still unstable, as expected.  We have verified that the
evolution in run E45N2b is indeed a result of baroclinic instability:
the perturbation fields tilt in the appropriate directions with
height, as seen in Fig.~\ref{fig8} for run E45N. The instability is,
however, weaker and evolves differently than when there is an initial
vertical wind shear and meridional entropy gradient at the lower
boundary.  We have performed a simulation with the initial flow
profile used in run E45N2b but with bottom raised to 1~bar, in which
the shear and gradient is non-zero at the bottom.  The peak global
eddy kinetic energy in this run is $\sim\!$~40 per cent greater and
vorticity perturbations are up to six times stronger than in run
E45N2b.  Nevertheless, the point is, the instability is present
regardless of vertical flow shear at the bottom boundary.

\subsection{High Latitude Westward Jet}\label{westward}

\begin{figure*}
  \vspace*{1cm}
  $\begin{array}{cc}
    \hspace*{-.3cm}
    \includegraphics[height=10.3cm]{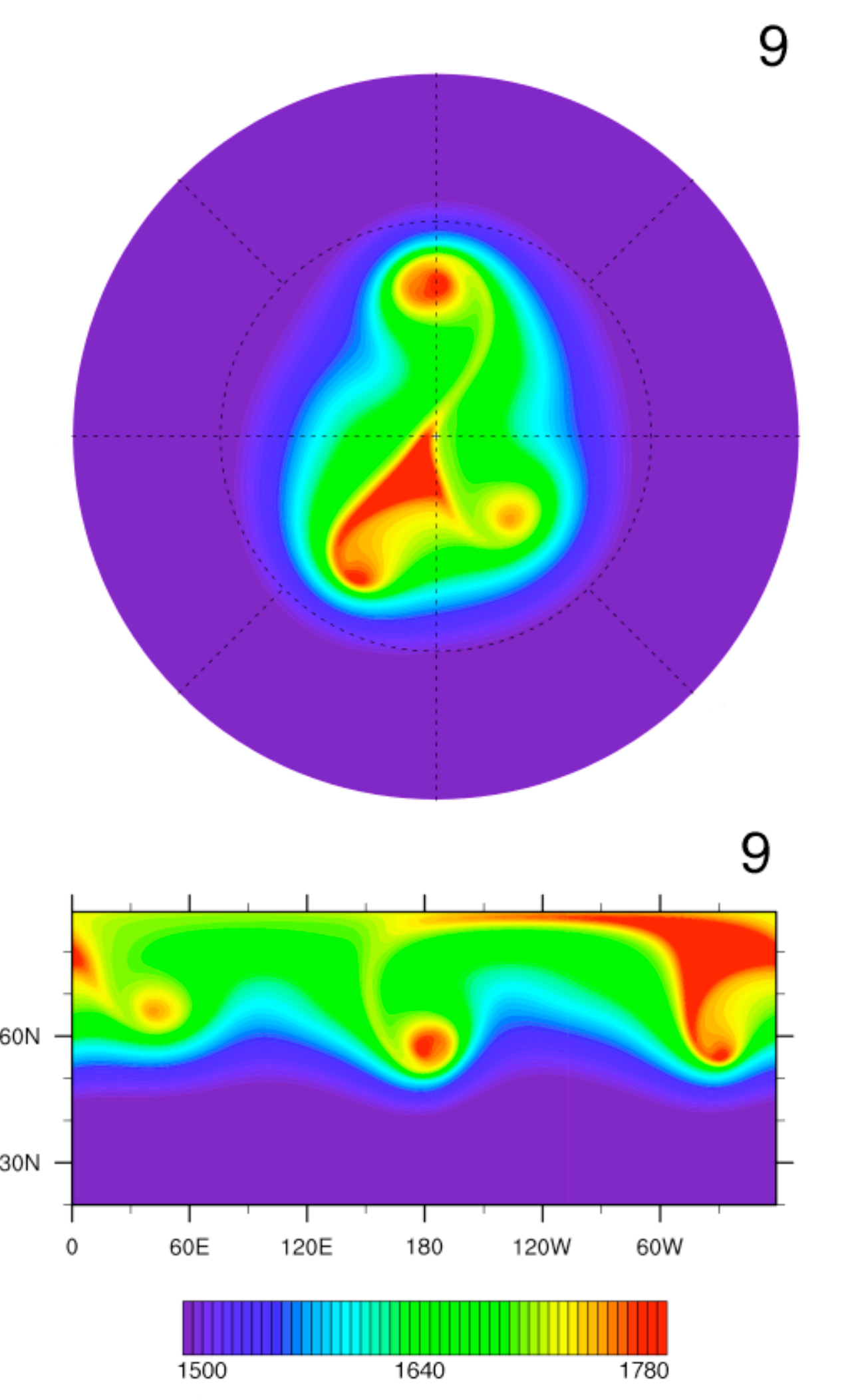}
    \includegraphics[height=10.3cm]{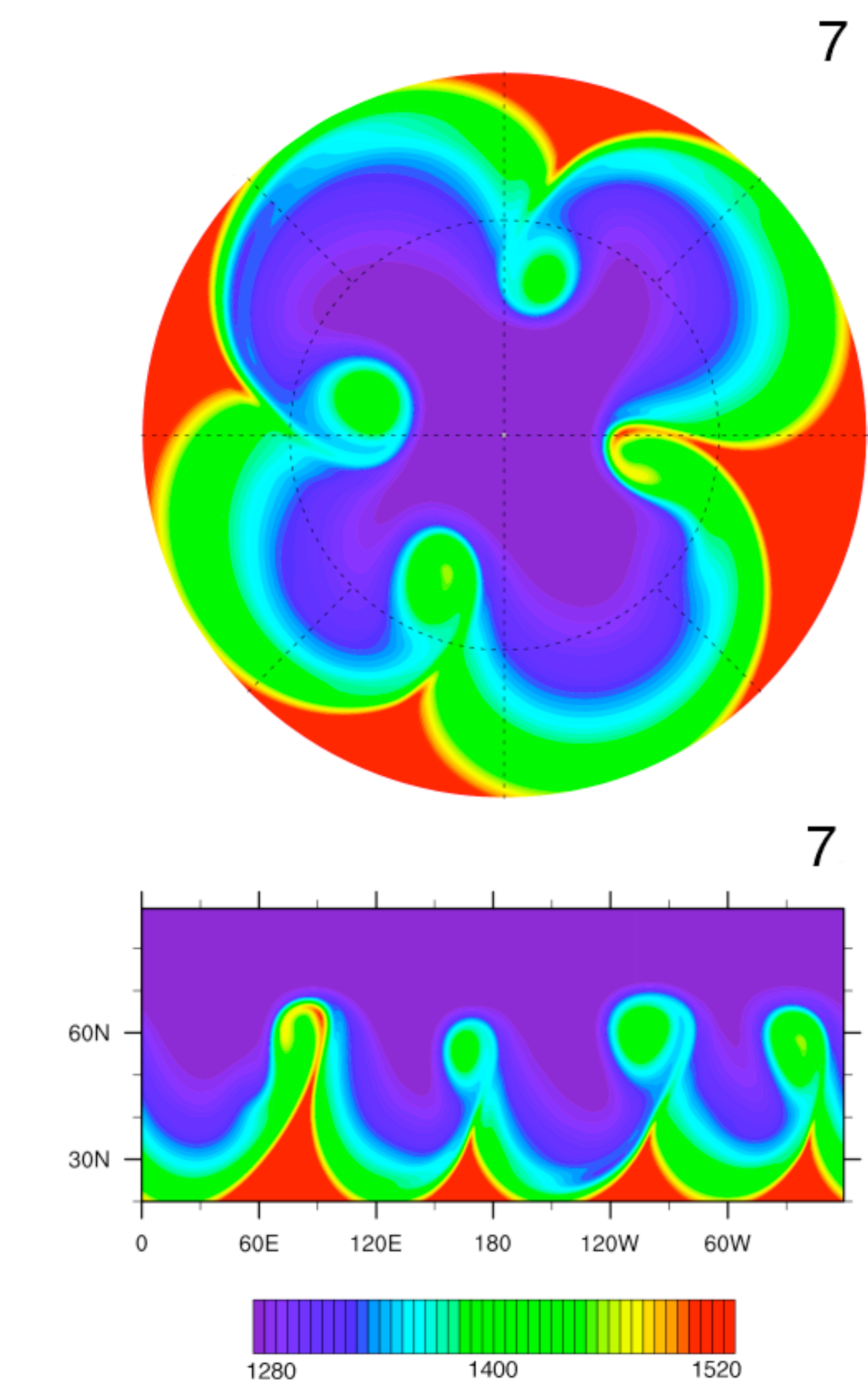}
  \end{array}$
  \vspace*{.3cm}
  \caption{Temperature field $T$ for run W60N at $\tau = 9$ (left) and
    $T$ for run E45N at $\tau = 7$ (right).  The fields are shown at
    975~hPa pressure level.  Top frame shows the field in polar
    stereographic view, centred on the north pole, and bottom frame
    shows the field in cylindrical-equidistant view, centred on the
    equator.  In all frames area poleward of $\phi = 20\degr$ is
    shown.  For run W60N the maximum and minimum values are 1500~K and
    1780~K, respectively.  For run E45N the maximum and minimum values
    are 1280~K and 1520~K, respectively.  Contour interval is 6~K in
    both runs.}
  \label{fig13}
\end{figure*}

Having presented the evolution of the `paradigm case', we now present
the case of baroclinically unstable high-speed westward jet at high
latitude (run W60N).  The speed at the core of the jet is
1000~m~s$^{-1}$. Such jets have been observed in recent GCM
simulations \citep[e.g.][]{Showman08,Menou09,Thrastarson10,Heng11}.
In these simulations, the high latitude jets tend to be more narrow
and shallow than the equatorial jets.  Equatorial jets will be
discussed in section~\ref{equatorial}.  The $T$ field from run W60N at
$\tau = 9$ (975~hPa pressure level) is shown in the left column of
Fig.~\ref{fig13}.  Polar stereographic view, centred on the north pole
(top frame), and cylindrical-equidistant view, centred on the equator
(bottom frame), are shown for latitudes poleward of $20\degr$N.  For
comparison, the right column shows the corresponding projections of
the $T$ field from the E45N run at $\tau = 7$, roughly at a similar
stage of the evolution in run W60N.

In run W60N, baroclinic wave develops a significant
northwest--southeast tilt.  This is consistent with the predominantly
negative momentum fluxes on the poleward side of the jet during the
linear stage of the evolution.  Again, the flow is characterized by
sharp cyclonic fronts, this time with the most unstable mode having 3
undulations at $\phi = 60\degr$.  The reduction in the number of
undulations is also consistent with linear theory developed in section
\ref{analysis}.  However, contrary to the predictions from the linear
analysis, the growth rate of the instability is lower than for the
E45N run (cf. the onset of growth between dashed and dotted lines in
Fig.~\ref{fig7}).  As already discussed, this agrees with the analysis
of the Charney model by \citet{Wang89} and extends that result to the
more general, global primitive equations model.

Note that wave-breaking in the westward jet case occurs in the
opposite direction to that in the `paradigm case' (see bottom row of
Fig.~\ref{fig13}).  The waves in run W60N (left) breaks eastward,
whereas in run E45N (right) the waves break westward.  'Blobs' of
higher temperature fluid penetrate into the lower temperature region
and cooler fluid subsides into the warm region.  The situation is
analogous to Rayleigh-Taylor or convective instability
\citep[e.g.][]{Sharp84}, where a decrease of potential energy results
under the interchange of two blobs at different heights.  In
baroclinic instability, this can occur despite the stable density
stratification because the density surfaces are sloping more steeply
than the line joining the two blobs.  Indeed, for this reason
baroclinic instability is sometimes refer to as `sloping convection'
\citep[e.g.][]{Vallis}.

Note that the \EKE for the westward jet case does not follow a simple
`baroclinic growth\ --\ barotropic decay' cycle, as seen in the
`paradigm case' (Fig.~\ref{fig7}).  Instead, after the initial decay
stage at $\tau\approx 15$, the {\it EKE} for run W60N shows large
vacillations, corresponding to a sequence of baroclinic--barotropic
life cycles.  Similar behavior of energetics has been observed by
\citet{Feldstein91} for the Earth case.  In a two-layer QG
$\beta$-plane model, he found westward jets to undergo a series of
mixed, baroclinic--barotropic instability, caused by the reversal of
sign in the jet curvature $\partial^2 u_0/\partial y^2$.  Recall that
$\beta \ge 0$.  Hence, a barotropically unstable region, in which
$\beta - \partial^2 u_0/\partial y^2 < 0$, forms at the core of the
westward jet (as is the case in run W60N).  The combined effects of
vertical and horizontal shears reinforce each other to establish a
mixed, baroclinic--barotropic unstable region.  According to WKB
analysis \citep[e.g.][]{Bender}, growing disturbances emanating from a
westward jet are trapped (i.e. reflected) between two turning
latitudes, initiating the sequence.  Consistent with this, the
meridional structure of the disturbance is able to remain close to the
normal mode form.  In contrast, disturbances emanating from the
eastward jet are absorbed at or near the critical latitudes, resulting
in a single cycle and meridional structure that changes with time.

Fig.~\ref{fig10}b shows the equilibrated $\bar{u}$ and $\bar{\theta}$
at the end of the simulation.  The original westward jet has been
completely disrupted, giving way to a fairly barotropic eastward jet
centred at 60$\degr$. Predominantly westward flow is now situated in
the subtropics, at the upper levels of atmosphere.  The reversal of
the flow direction is consistent with E-P flux divergence shown in
Fig.~\ref{fig11} (dotted line), which acts as a positive momentum
source. Given that high-latitude westward jets appear to be a fairly
common feature in GCM simulations of hot giant extrasolar planets, the
result here suggests external (e.g. stellar irradiation) or internal
(e.g. wave) forcing may be required to maintain baroclinic westward
jets.  Note also from Figs.~\ref{fig10}b and \ref{fig11}, the negative
zonal flow and E-P flux convergence, especially in the equatorial
region.  Significantly, such negative E-P flux divergences present a
source of drag for equatorial jets.  Finally, as in the `paradigm
case', the potential vorticity anomalies exceed $2\Omega$ by over 12
per cent and by $\tau = 35$ a cyclonic polar vortex forms that is
warmer than its surroundings (not shown).

\subsection{Equatorial Jet}\label{equatorial}
\begin{figure*}
  \vspace*{.5cm}
  \includegraphics[scale=.73]{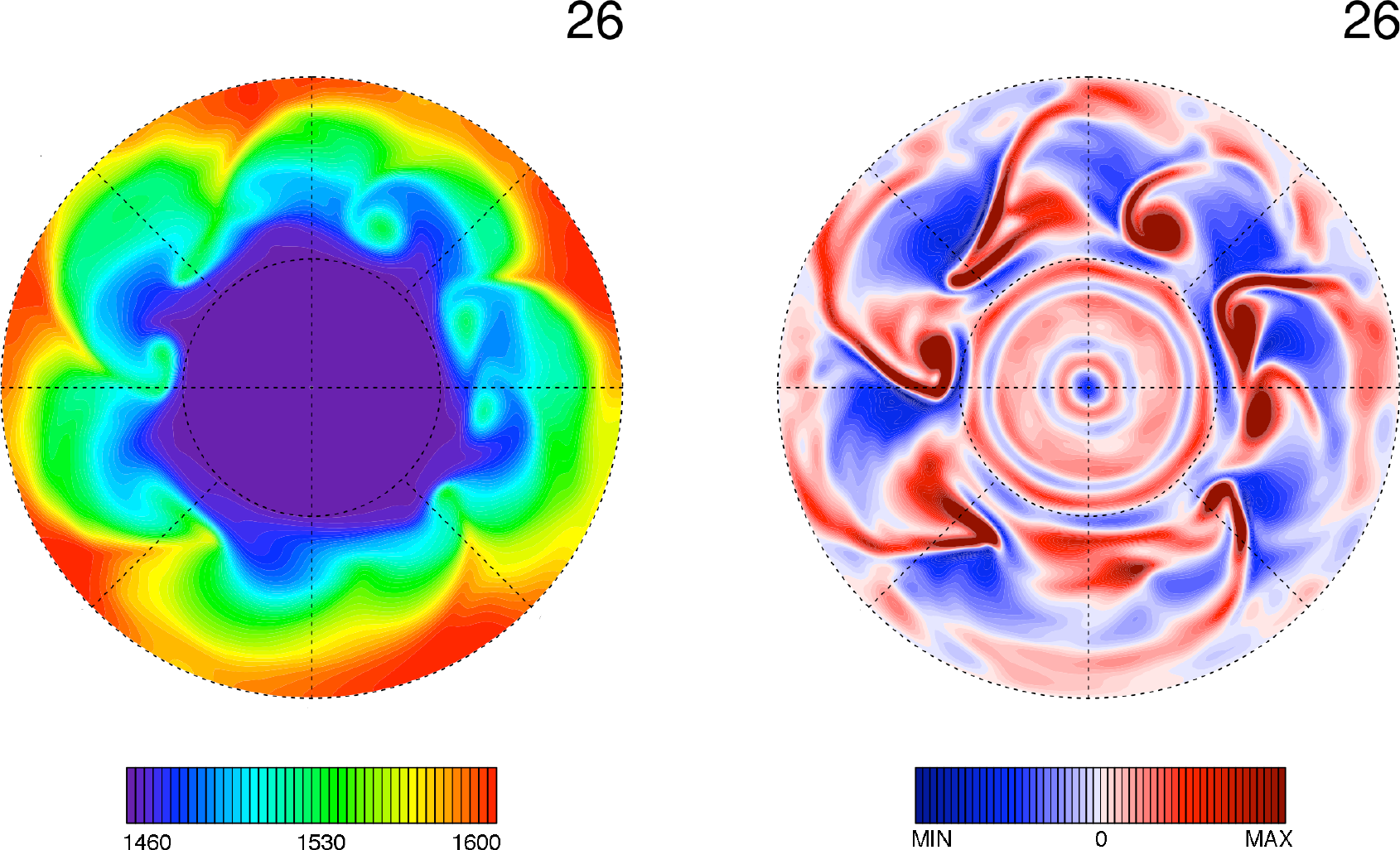}
  \vspace*{.3cm}
  \caption{Temperature $T$ (left) and relative vorticity $\zeta$
    (right) for run EEQ in polar stereographic view, centred on the
    north pole.  The fields are shown at 975~hPa pressure level for
    $\tau = 26$.  Maximum and minimum values for temperature are
    1460~K and 1600~K, respectively, with contour interval 3.5~K.
    Values for $\zeta$ are in the range $\,\pm 1.5 \times
    10^{-5}$s$^{-1}$, with contour interval $6 \times
    10^{-7}$s$^{-1}$.}
  \label{fig14}
\end{figure*}

The evolution of a broad, high-speed equatorial jet (run EEQ) is
presented in this section.  The initial flow and potential temperature
is shown in Fig.~\ref{fig4}c.  Unlike the jets discussed in
sections~\ref{base case} and \ref{westward}, the equatorial jet
satisfies the Charney-Stern-Pedlosky instability criteria (iv) on its
flanks (at $\sim\,$30$\degr$ on both northern and southern
hemispheres), rather than at the core.  The stability of this jet's
core is consistent with linear theory of section~\ref{linear}, which
predicts no growth for a jet located equatorward of $28\degr$.

Fig.~\ref{fig14} shows temperature $T$ and relative vorticity $\zeta$
fields at $\tau = 26$.  At this time the instability is well
developed, with sharp fronts rolling up non-linearly into cyclones at
$\phi\approx 35\degr$, where the instability criteria is met.  A mode
with $\sim\! 7$ undulations can clearly be seen at this stage of the
evolution.  The number of undulations is significantly higher and the
growth rate is significantly lower for EEQ than for simulations where
the same jet is placed at $\phi = 30\degr$.  Evidently, since the
vertical shear of the equatorial jet at its flanks is significantly
lower than at its core, a value smaller than the peak core value for
the characteristic flow speed should be used in interpreting the
results from the linear analysis.  We have already seen that a weaker
jet (shear) results in a smaller growth rate and wavelength of the
most unstable mode at a given latitude (e.g. curves `HD45L' and `HD45'
in Fig.~\ref{fig2}).  Hence, our non-linear calculations appear to be
in qualitative agreement with linear theory.  Despite the instability
at the flanks, the core of the jet in run EEQ remain stable throughout
the integration (up to $\tau = 60$), in very good agreement with
linear theory.

\begin{figure*}
  \vspace*{.5cm}
  \includegraphics[width=0.59\textwidth]{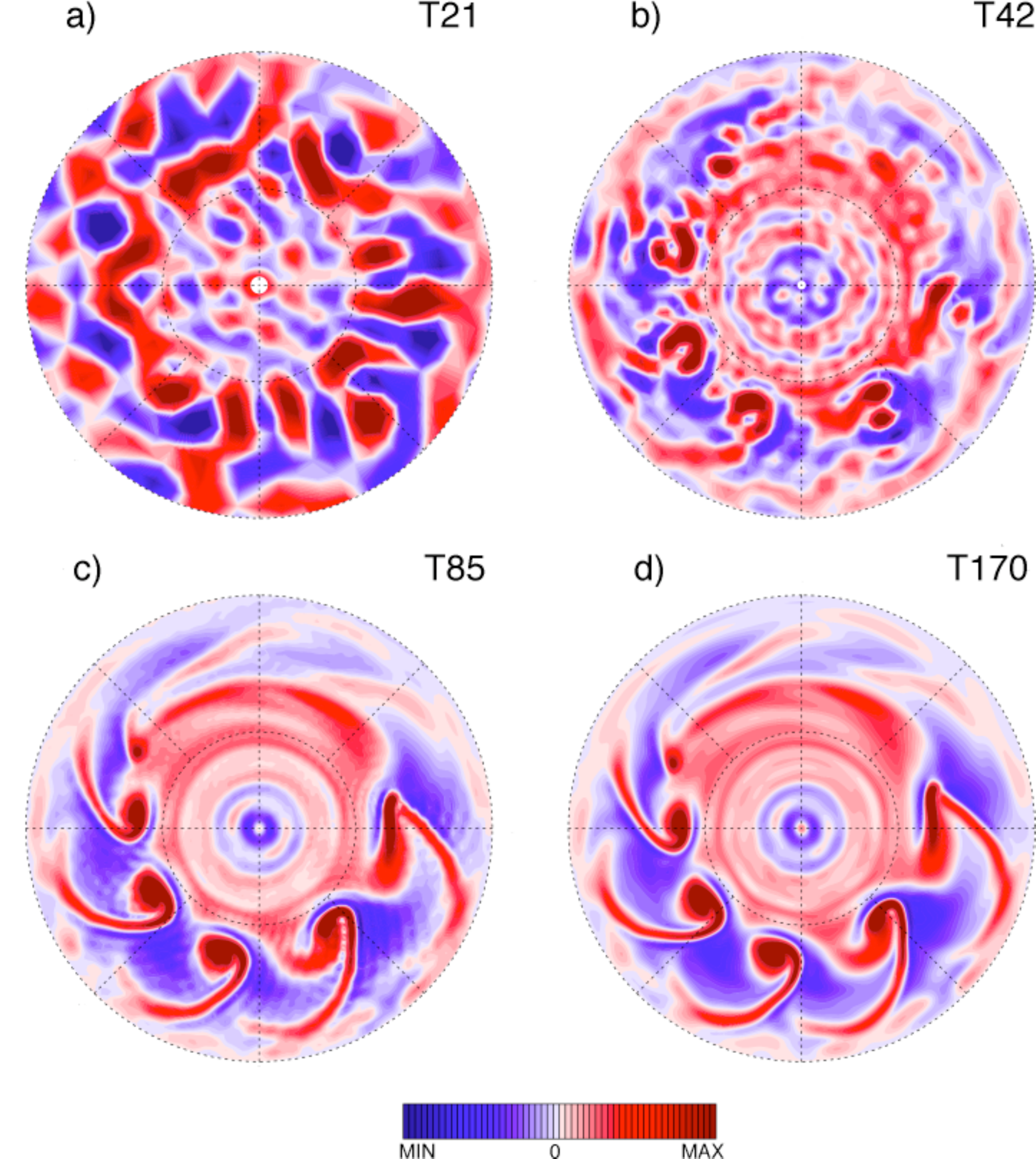}
  \vspace*{.4cm}
  \caption{Polar stereographic view of the relative vorticity field
    $\zeta$, centred on the north pole, for four runs with all
    parameters identical -- except the horizontal resolution.  The
    common parameters are as in run EEQ.  The fields at $\tau = 22$
    are shown.  The number at upper right in each panel indicates the
    resolution.  Contour levels are the same to those in
    Fig.~\ref{fig14}.}
  \label{fig15}
\end{figure*}

The {\it EKE} evolution for run EEQ is shown in Fig. \ref{fig7} (solid
line).  The equatorial jet instability is shallow and confined to a
pressure range between 1 to 0.7~bar, unlike the high-latitude jet
instability; in those cases, the range of instability is much larger,
extending up to 0.01~bar.  Thus, only the lower pressure levels
exhibit an increase in {\it EKE} during the linear stage.  For this
reason, the {\it EKE} values have been multiplied by a factor of 50 in
the figure: the globally averaged {\it EKE} for run EEQ is much lower
than for the `paradigm case' or run W60N.  Qualitatively, the
non-linear evolution of run EEQ is much like that of run E45N, with
waves tilting and breaking in same directions.  However, potential
vorticity anomaly only slightly exceeds the polar value and the
cyclonic drift does not ensue.  Therefore, a monolithic polar vortex
does not form in this case.

Interestingly, the jet structure is only slightly altered by
baroclinic instability from the basic state zonal flow.  Mainly, the
jet has become more barotropic at the flanks.  This can be seen in
Fig.~\ref{fig10}c, which shows $\bar{u}$ and $\bar{\theta}$ at $\tau =
60$. Relatively small values of E-P flux convergence equatorward of
$30\degr$ (see the solid line in Fig.~\ref{fig11}) do not
significantly contribute to the deceleration of the zonal mean zonal
wind.

It is also worth noting that we have investigated stability properties
of the westward equatorial jet and found it to be stable to baroclinic
instability, in good agreement with \citet{Wang89}.  A westward jet
placed at the equator would have to exceed sound speed, if the
condition (\ref{crit_shear}) of section~\ref{linear} is to be
fulfilled.  However, we note that a broad, `supersonic' westward jet
does not appear to be unstable in full, non-linear GCM calculations
(Thrastarson, private communication).  But, the calculation is at T21
resolution (see next section).  We have found eastward equatorial jets
to be stable to baroclinic instability, if the width of the jet is
50$\degr$ and smaller.

\begin{figure*}
  \hspace*{-.5cm}
  $\begin{array}{cc}
    \includegraphics[height=7.5cm,width=8.9cm]{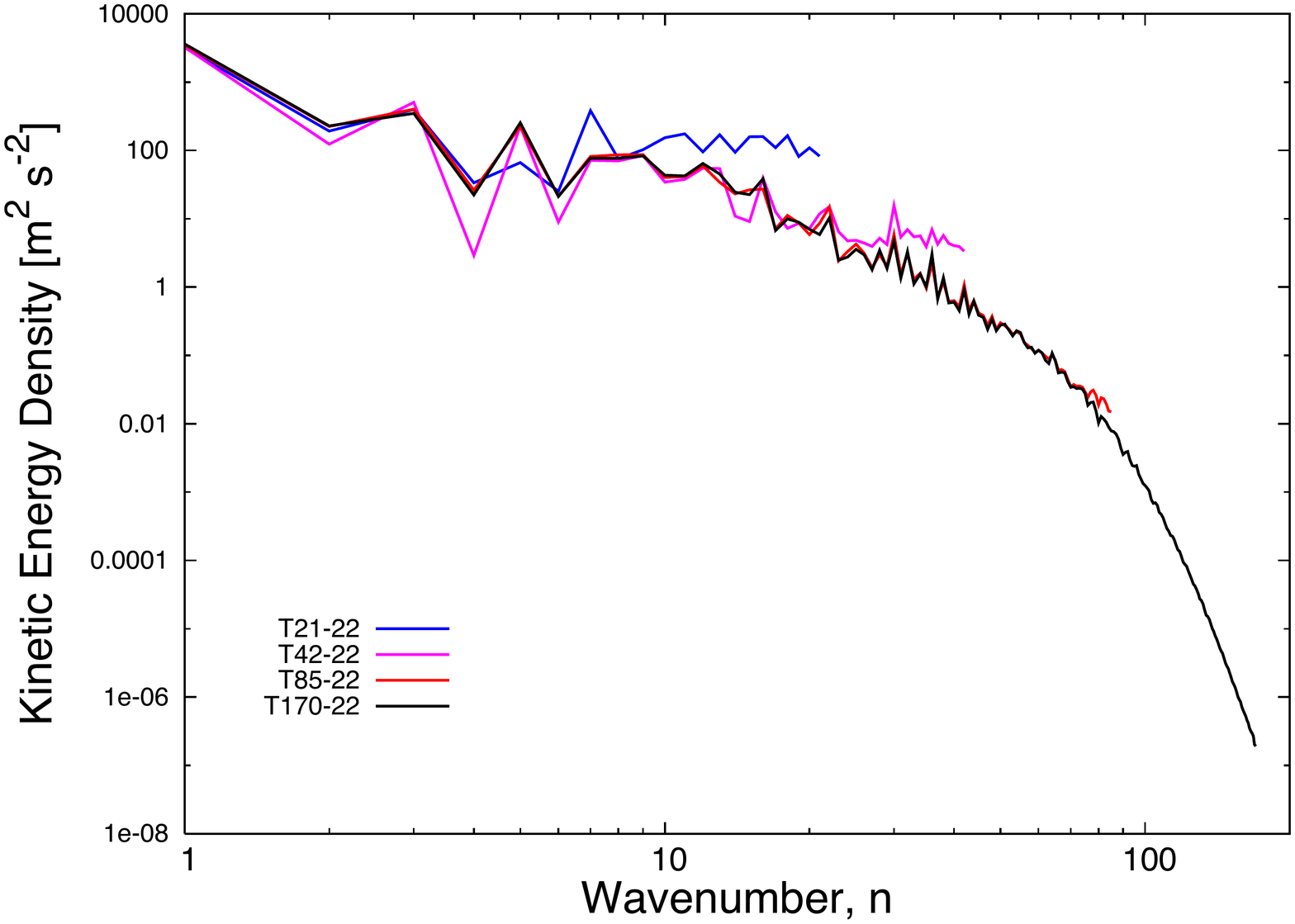}
    \hspace*{.3cm}
    \includegraphics[height=7.5cm,width=8.9cm]{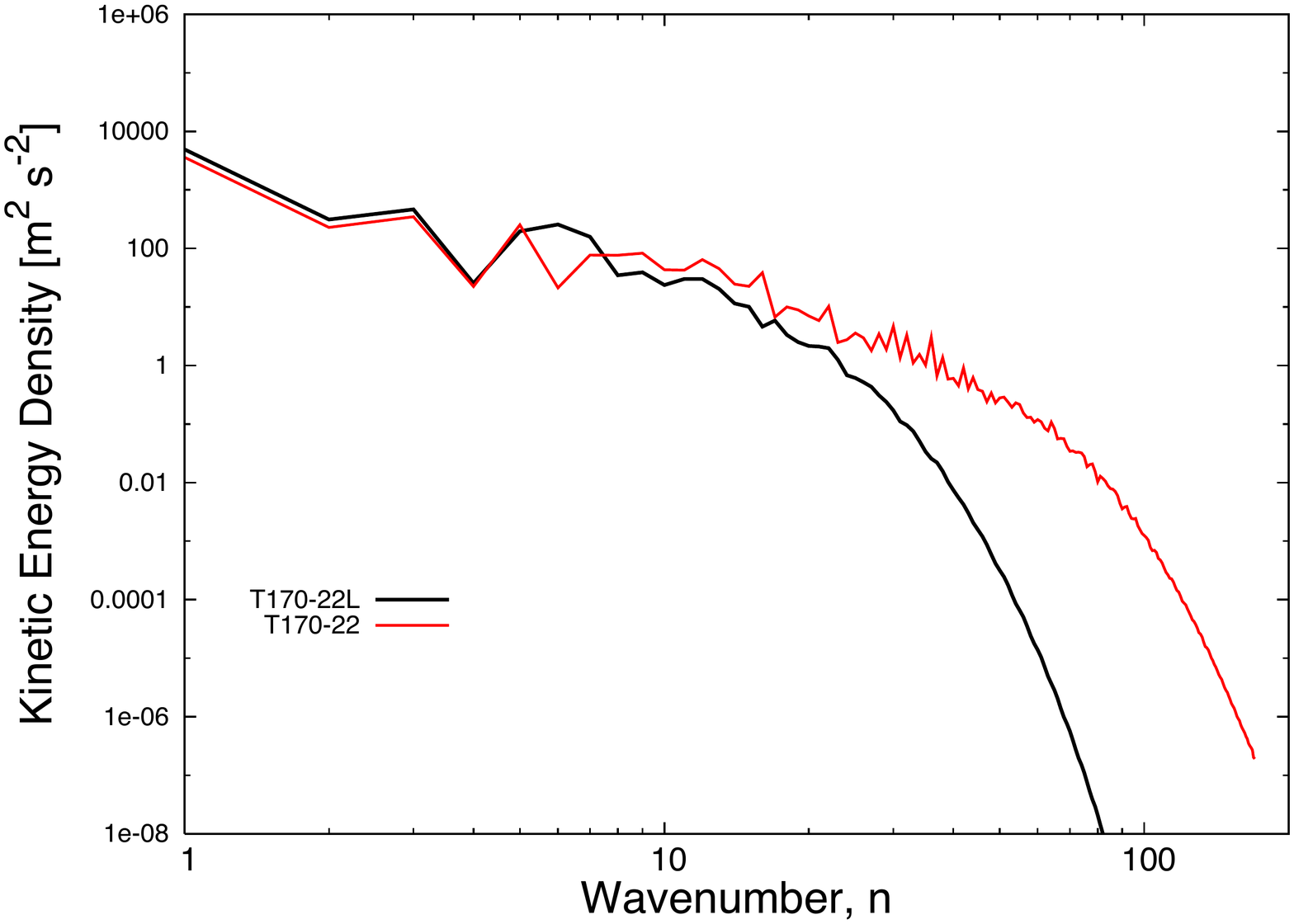}
  \end{array}$
  \vspace*{-.5cm}
  \caption{Kinetic energy density [m$^{2}$~s$^{-2}$ (per wave number)]
    as a function of (total) wavenumber $n$.  Spectra for the fields
    from the four runs shown in Fig.~\ref{fig15} (left).  The
    different lines refer to different horizontal resolutions, as
    indicated in the legend.  The viscosity coefficient is same ($\nu
    = 6\times 10^{19}$ m$^4$ s$^{-1}$) in all four runs.  Spectra for
    run EEQ at T170 resolution with different viscosity
    coefficients~(right): $\nu = 6\times 10^{19}$ m$^4$ s$^{-1}$ (red
    line) and $\nu= 10^{21}$ m$^4$ s$^{-1}$ (black line),
    respectively.}
    \label{fig16}
\end{figure*}

\subsection{Numerical Convergence}\label{convergence}
\begin{figure*}
  \includegraphics[width=0.59\textwidth]{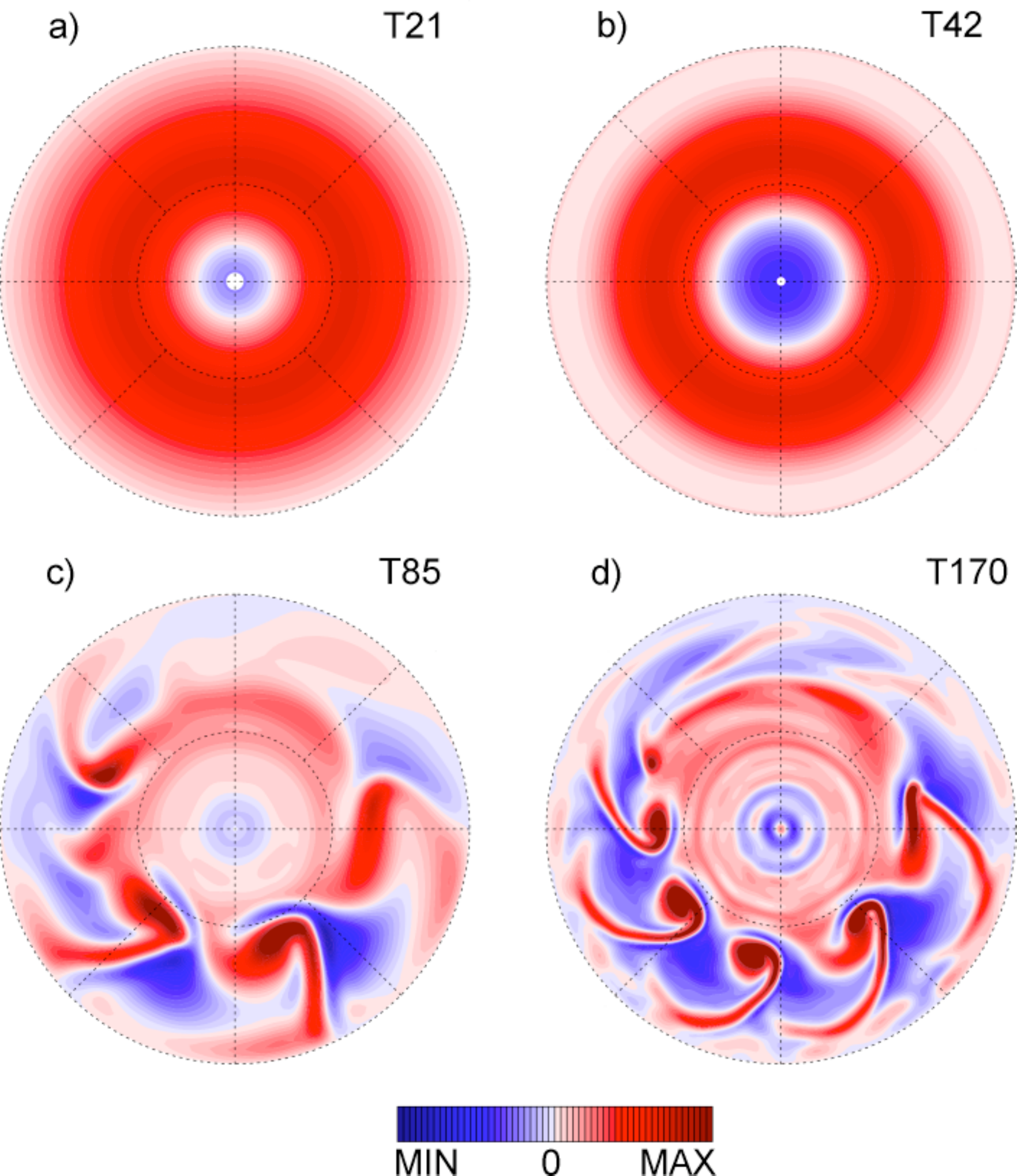}
  \caption{Same as Fig.~\ref{fig15} but with $\tau_{\rm d}^{-1} =
    5.07\times 10^{-4}$~s$^{-1} $ in all runs.  Maximum and minimum
    values for frames a) and b) are $\pm 3 \times 10^{-6}$~s$^{-1}$
    and for frames c) and d) $\pm 1.5 \times 10^{-5}$~s$^{-1}$; the
    contour intervals are, respectively, $1.4 \times 10^{-7}$~s$^{-1}$
    and $6 \times 10^{-7}$~s$^{-1}$.}
  \label{fig17}
\end{figure*}

Baroclinic instability in numerical simulations of hot extrasolar
planets is highly sensitive to numerical resolution (both horizontal
and vertical) and to dissipation.  High resolution is required to
capture the instability accurately.  In particular, for the jet
profiles used, five or more layers is necessary to capture the
instability, and good convergence is reached only with $\sim\! 10$ or
more layers.  In addition, high horizontal resolution (\,$\ga$~T85) is
necessary to ensure accurate representation of the eddy fluxes, as
well as convergence.  Separately, artificial viscosity must not be too
high, as it results in artificially-enhanced stabilization of the
baroclinic modes.  We emphasise that resolution and dissipation
requirements are dependent on the jet profile.  Hence, the
requirements should be carefully assessed for each profile employed.
This `problem-dependence' conclusion has also been discussed by
\citet{Thrastarson11} for `spin-up' experiments of hot giant planet
circulations.

Before presenting the results, a brief discussion concerning our
general approach to convergence testing is in order.  In general, for
numerical stability reasons, the usual practice is to use a larger
dissipation coefficient value when performing a calculation with lower
resolution -- or, alternatively, a smaller coefficient value when
performing the same calculation at higher resolution -- so that the
damping time is same at the truncation scale.  This results in a
different damping time for a given mode at different resolutions.
However, here our aim is to demonstrate convergence of the numerical
model.  Hence, we employ the same value of the coefficient at all
resolutions so that each mode, up to the truncation, experiences same
dissipation rate in all the runs.  The employed value is: $\nu =
6\times 10^{19}$~m$^4$~s$^{-1}$.  Similar methodology has been
implemented in e.g. \citet{Polvani04} to test convergence in the Earth
case.  Later, we also demonstrate non-convergence when the damping
time is chosen such that it is same at the truncation scale for all
resolutions, as in the usual practice.

The requirement of adequate resolution is demonstrated in
Fig.~\ref{fig15}.  The figure shows a set of four simulations with all
parameters identical to run EEQ presented in section~\ref{equatorial}
-- except the resolution.  The resolutions are: T21~(a), T42~(b),
T85~(c) and T170~(d).  Note, panel (d) is run EEQ.  All four runs use
the same value of superviscosity coefficient, $\nu = 6\times
10^{19}$~m$^4$ s$^{-1}$, as already mentioned.  Note also that the
resolutions correspond, respectively, to 64$\times$32, 128$\times$64,
256$\times$128 and 512$\times$256 Gaussian grids in physical space.
But, because of the exponential convergence property of the spectral
method, they are equivalent in accuracy to finite difference grids
${\cal O}(10)$ times finer in resolution
\citep[e.g.][]{Thrastarson11}.  Polar stereographic projections of the
relative vorticity field $\zeta$ at $\tau = 22$ are shown, when the
instability is in the early exponentially-growing stage (see
Fig.~\ref{fig7}).

Visual inspection of the fields readily reveals that the T21~(a) and
T42~(b) runs do not converge to the T170~(d) run.  The T85~(c) run is
marginally converged, though this may change after a long time (e.g.
many hundreds of planetary rotations).  In the figure, frames~(a)
and~(b) are qualitatively different than frames (c) and (d), which
clearly show mode-6 instability.  The T85 run in frame~(c) captures
the basic structure present in the T170 run in frame~(d).  However,
spurious small-scale oscillations are also clearly visible in
frame~(c); these are not present in frame~(d).  The small-scale
oscillations contaminate the calculation -- causing the calculation to
blow up, depending on the numerical parameters used; see e.g.
discussion in \citet{Thrastarson11}.

The above behavior can be quantified by computing the corresponding
kinetic energy spectra for each run.  The spectra for the fields shown
in Fig.~\ref{fig15} are presented in the left panel of
Fig.~\ref{fig16}.  Inspection of the T85 and T170 spectra (red and
black lines, respectively) confirms the convergence of the
simulations.  Note the presence of a clear dissipation range in the
T170 run.  In contrast, the appearance of nearly grid-scale waves in
physical space for T21 and T42 resolution runs corroborates the
tendency of the spectrum (blue and pink lines for T21 and T42,
respectively) in these runs to peel off and curl up considerably left
of the aliasing limit ($\sim\,$21 for T21 and $\sim\,$42 for T42).
This is caused by discretization errors that are not adequately
controlled by the applied explicit viscosity.

We have also performed an analogous series of runs in which a much
larger dissipation coefficient value, $\nu = 10^{21}$~m$^4$~s$^{-1}$
has been used.  This mimics `properly' dissipated runs at T21 and T42
resolutions (i.e. runs with well represented dissipation range in
spectral space).  However, in this series the high resolution
calculations are significantly over-dissipated and the physical space
picture is characterised by a severe reduction in eddy kinetic energy
at all times.  This is supported by the spectra for two T170
resolution simulations with the two coefficients (right frame of
Fig.~\ref{fig16}).  The spectrum for a run using $\nu =
10^{21}$~m$^4$~s$^{-1}$ (black line) is shown together with the
spectrum of the previously presented T170 run with $\nu = 6\times
10^{19}$~m$^4$~s$^{-1}$ (red line).  With the larger $\nu$, the
spectrum is severely over-dissipated with only $\sim\!  20$ modes
being resolved; the rest of the modes clearly lie in the dissipation
range.  In contrast, at least $\sim\!  80$ modes are well-resolved
with the smaller $\nu$ value.

It is important to understand that wavenumbers short-ward of the
fiducial `dissipation range' (i.e. less than 20 and 80, respectively,
in the runs discussed above) are still affected by a small amount of
dissipation in practice: that is, dissipation affects the entire
spectrum of wavenumbers continuously, rather than just the wavenumbers
in the dissipation range.  The amount, while small, can nevertheless
be {\it dynamically} significant, as it can change the quantitative
character of the instability, even suppress the instability
altogether.  Indeed, if the value of $\nu$ is increased further, to
$10^{22}$~m$^4$~s$^{-1}$, the baroclinic waves completely disappear.

As discussed above, a common practice in numerical studies which vary
the resolution is to adjust the dissipation coefficient $\nu$ so that
the damping time $\tau_{\rm d}$ is same at the truncation scale $n_t$:
  \begin{equation}\label{tauD}
    \tau_{\rm d} = 
    \frac{1}{\nu}\left[ \frac{R_p^2}{n_t (n_t + 1)}\right ]^2.
  \end{equation}
  While the physical basis of this procedure is arguable, we
  demonstrate here that the practice still does not lead to
  convergence at low resolution.  Consider the value, $\nu = 6\times
  10^{19}$~m$^4$~s$^{-1}$, used in the high resolution simulation
  discussed above.  The dissipation rate at the truncation scale for
  HD209458b corresponding to this $\nu$ value is $\tau_{\rm d}^{-1} =
  5.07\times 10^{-4}$~s$^{-1} $.  Note that this damping rate is
  comparable to the rates used in current flow modeling studies of hot
  gaseous extrasolar planet atmospheres at resolutions lower than T170
  \citep[e.g.][]{Rauscher10,Thrastarson10,Thrastarson11,Heng11}.  Its
  significance can be seen in Fig.~\ref{fig17}.

  Fig.~\ref{fig17} shows the effects of adjusting $\nu$ so that the
  damping rate is constant at the truncation scale.  The rate used is
  the one just discussed above: $\tau_{\rm d}^{-1} = 5.07\times
  10^{-4}$~s$^{-1}$.  Relative vorticity field $\zeta$ at four
  resolutions for run EEQ at $\tau = 22$ is shown.  Two points are
  clear from the figure.  First, the simulations in this series are
  not converged.  The T21 and T42 resolution runs are completely
  over-dissipated and the momentum and heat transports are absent
  throughout the duration of the runs, up to $\tau = 60$.  Second, the
  $\nu$ values used in current GCM modelling studies of hot extrasolar
  giant planets do not permit the instability.  Note that, if the
  dissipation rate is chosen instead to be the one that `adequately'
  permits the instability in the low resolution run, the high
  resolution runs are severely under-dissipated and inundated with
  noise (not shown).  Either way, convergence is not achieved by
  fixing the dissipation rate at the truncation scale.

  Arguably, the two points above may not be significant for
  atmospheres characterized by a very short diabatic relaxation time.
  For then the thermal damping would dominate and naturally
  short-circuit the above issues pertaining to the artificial
  viscosity and resolution.  However, in some GCM simulations, the
  dynamically-relevant, intrinsic thermal relaxation time is not
  always short after quasi-equilibration is reached, even above the
  $\sim$1~bar level\footnote{It is, in general, sufficiently long
    below this level.} (Thrastarson (private communication)).
  Moreover, the instabilities at higher latitudes possess short growth
  times and are much less affected by short relaxation times.
  Additionally, there is the issue of transient growth, which we have
  not discussed in this work.  The non-normal modes, associated with
  such growth, may operate on a much shorter time scale than the
  growths described in this work.

  We stress here that the high resolution runs described do not merely
  contain more fine-scale structures that presumably do not
  significantly affect the evolution.  On the contrary, we have found
  that high resolution fundamentally affects the evolution.  For
  example, bulk eddy heat- and momentum-fluxes differ significantly
  (up to an order of magnitude) in high- and low-resolution
  simulations.  The reason for this is that vorticity anomalies
  (eddies) are much stronger, in addition to the filaments, in the
  high-resolution simulations; and, \EKE growth is exponential in the
  stage of the evolution when these structures have emerged
  (Fig.~\ref{fig7}).  Hence, heat and momentum are redistributed much
  more effectively.

\section{Summary and Discussion}\label{summary}

Baroclinic instability on extrasolar planets has not been studied thus
far.  In this work we have used an advanced pseudospectral GCM to
perform an extensive study of the stability and non-linear evolution
of balanced jets on hot extrasolar planets.  Our non-linear baroclinic
instability calculations have been fully validated against previous
similar calculations for the Earth
\citep[e.g.][]{Polvani04,Jablonowski06}.  For concreteness, we have
presented here results for a model planet with physical parameters
corresponding to the close-in giant planet HD209458b and focused on
the stability of high-speed (typically 1000~m\,s$^{-1}$) eastward jets
at the equator and westward jets at high latitudes.  Broad jets of
such magnitude are a common feature in current GCM simulations of
tidally-synchronized giant planet atmospheres.

We have derived linear growth rate and phase speed spectra, via
standard normal mode analysis, and compared the results with full
non-linear numerical simulations.  According to our linear analysis of
the two-layer primitive equations model on the $\beta$-plane, the
growth rate of the instability is reduced for a jet located at low
latitudes, compared with a jet located at high latitudes.  Near the
equator, where the deformation length scale $\LD$ becomes too large to
accommodate baroclinic waves, the linear theory predicts stability.
In general, linear analysis agrees reasonably well with the full
non-linear calculations at the early stage of the unstable evolution,
during the transient phase.  After a long time, in simulations with
high values of initial potential vorticity anomaly (i.e. $|q^\prime /
2\Omega| \ga 1.2$, where $q^\prime$ is the anomaly), cyclones merge to
form robust monolithic vortices at the poles. This is not captured by
linear analysis.

As expected, full non-linear calculations show richer behavior than
that obtained through linear analysis.  Non-linear simulations show
that baroclinic instability occurs for all eastward jet profiles used
in this study.  In particular, broad equatorial eastward jets are
unstable (on a time scale of $\sim$\,20~planetary rotations), despite
stability suggested by the linear analysis.  The instability takes
place at the jet flanks, where there is still a significant vertical
shear to satisfy the necessary condition for instability.  The jet
core is stable, unlike in the jets situated at higher latitudes; this
is in accordance with linear theory. Westward jets near the equator,
however, remain stable, both at the core and the flanks.  To the best
of our knowledge this is the first time non-linear baroclinic
instability has been studied for a broad equatorial jet in the
atmosphere.  In general, we have found westward jets to be more stable
compared to their eastward counterparts (e.g. at midlatitude,
instability timescale of $\sim$\,6~planetary rotations for westward
jets vs. $\sim$\,3~planetary rotations for eastward jets), and to
require much stronger vertical shear for instability in the full
primitive equations system.  Additionally, we have demonstrated in
this work that baroclinic instability does not require a solid
boundary on planets, as long as there is a change of sign in the
meridional potential vorticity gradient $\partial q_0/\partial\phi$ in
the domain's interior.

By performing the simulations described above with a wide range of
horizontal resolution (from T21 to T170), we have found that the
calculations do not converge for resolutions below T85.  This is a
somewhat stronger requirement than for Earth simulations and is
primarily due to the much stronger jet amplitude on hot extrasolar
planets ($\sim\,$1000~m~s$^{-1}$, compared to $\sim\,$50~m~s$^{-1}$
for the Earth).  Furthermore, we have found that baroclinic
instability does not occur at all if the artificial viscosity
coefficient used in the calculation is too high.  A high artificial
viscosity is often used to stabilize numerical simulations against
strong forcing in current studies of extrasolar planet atmospheres.
Given this, baroclinic instability is unlikely to be represented in
current simulations -- even when necessary conditions for instability
are satisfied. This may pose a serious issue in flow modelling studies
of extrasolar planet atmospheres in which the natural diabatic
relaxation time is not too short (i.e. greater than a few planetary
rotations).

The results presented in this paper show that baroclinic instability
is significant for understanding characteristics of hot extrasolar
planets which possess fast jet streams.  This instability is likely to
play a role in weather, general circulation and large scale
variability of a few to few tens of planetary rotation periods on
these planets \citep{Cho08b}.  For example, the process could generate
large long-lived storms that could be observed remotely.  Sharp fronts
produced in baroclinic instability life-cycles can also act as a
gravity wave source \citep[e.g.][]{Osullivan95,Plougonven07}.  Gravity
waves are expected to play an important role in stably stratified
atmospheres of hot extrasolar planets: they can modify the circulation
through exerting accelerations (positive and negative) on the mean
flow, as well as transporting heat vertically from deep regions to
sensible regions and laterally from day side to night side
\citep{Watkins10}.

In this work, we only discuss adiabatic calculations.  Adiabatic
calculations are important to cleanly delineate many subtle effects in
rotating-stratified fluid that could obscure baroclinic instability.
They are also important as a foundation for the instability under
forced conditions, which require careful study.  The complex effects
of forcing on the background flow itself remains to be elucidated.  We
have also focused mainly on the instability and subsequent evolution
of jets in isolation and only lightly touched on the effect of
concomitant eddies on the background flow.  In summary, the full
effect of baroclinic instability on the mean flow on hot extrasolar
planets remains to be carefully studied.  Some of the issues
identified here will be addressed in future work.

\section*{Acknowledgments}

The authors thank Heidar Thrastarson and Chris Watkins for useful
discussions and the reviewer for helpful comments.  This work is
supported by a research studentship from the Westfield Trust to
I.P. and the Science and Technology Facilities Council grant
PP/E001858/1 and the Westfield Small Grant to J.Y-K.C.
\vspace*{1cm}

\appendix
\section{Non-dimensional Stability Analysis}
\begin{figure*} 
  \vspace*{.5cm}
  \includegraphics[height=6.1cm,width=17.5cm]{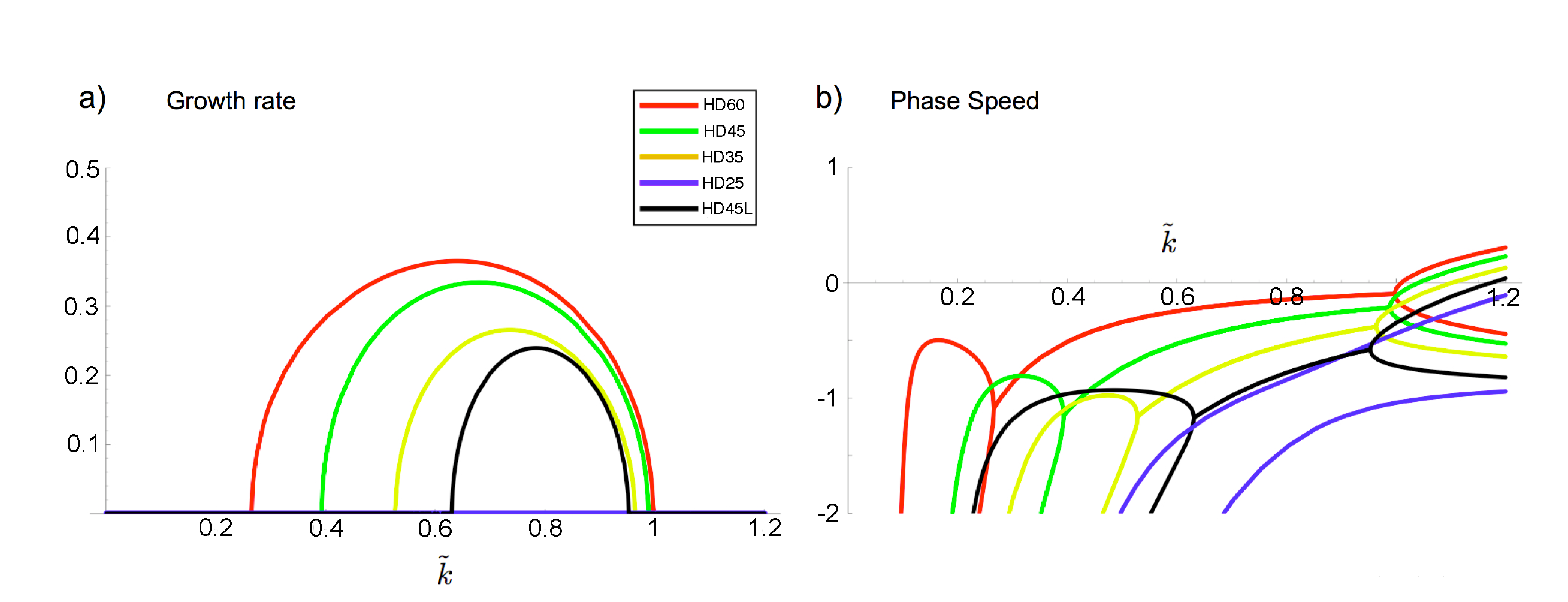}
  \vspace*{.3cm}
  \caption{Non-dimensional growth rate [$\tilde{k} \cdot
    \Im\mathfrak{m}\{\tilde{c}\}$] (left) and phase speed
    [$\Re\mathfrak{e}\{\tilde{c}\}$] (right) for HD209458b, as a
    function of non-dimensional wavenumber $\tilde{k}$.  Curves
    `HD60', `HD45', `HD35' and `HD25' represent growth rates and phase
    speeds at $\phi = (60\degr, 45\degr, 35\degr, 25\degr)$ computed
    with ($\hat{Ro}$= 0.76, $\hat{\gamma}$ = 0.14), ($\hat{Ro}$= 0.76,
    $\hat{\gamma}$ = 0.3), ($\hat{Ro}$= 0.76, $\hat{\gamma}$ = 0.51),
    ($\hat{Ro}$= 0.76, $\hat{\gamma}$ = 1.04). Curve `HD45L' has been
    computed for HD209458b parameters at $\phi = 45\degr$, but with
    $U_0 = 200$~m~s$^{-1}$ corresponding to ($\hat{Ro}$= 0.3,
    $\hat{\gamma}$ = 0.72). To obtain dimensional values, multiply the
    growth rate by $U_0 \sqrt{2}/\hat{L}_{\rm{\tiny D}}$ and wavenumber by
    $\sqrt{2}/\hat{L}_{\rm{\tiny D}}$.}
  \label{figA1}
  \vspace*{.3cm}
\end{figure*}

Equations set (\ref{Dimensional system}) is made non-dimensional by
introducing the `discretized deformation length scale', horizontal
length scale, timescale, and height scale:
\begin{eqnarray*}
  \hat{L}_{\rm{\tiny D}} & = & 
  \frac{1}{f_0}\,\sqrt{h_2 \triangle p \sigma_0}\, , \\
  L\  &  = &  \frac{1}{\sqrt{2}}\,\hat{L}_{\rm{\tiny D}} \, , \\
  T\  & = & 
  \frac{1}{\sqrt{2}}\left(\hat{L}_{\rm{\tiny D}} / U_0\right)\, , \\ 
  H\  & = & 
  \frac{1}{\sqrt{2}}\left(U_0 f_0\hat{L}_{\rm{\tiny D}}\right)\, ,
\end{eqnarray*}
respectively.  Hence, the set of equations becomes characterised only
by the Rossby number,
\[
\hat{Ro} = \frac{U_0 \sqrt{2}}{f_0 \hat{L}_{\rm{\tiny D}}}\, ,
\] 
and the Charney-Green number, 
\[
\hat{\gamma} = \frac{\beta\,\hat{L}_{\rm{\tiny D}}^2}{2\,U_0}\, .
\] 
The Charney-Green number measures the relative importance of the
planetary vorticity gradient to the relative vorticity gradient.
Using these non-dimensionalised scales and parameters, the equations
now read:
\begin{subequations}\label{perturbed eqns}
\begin{eqnarray}
  \frac{\partial}{\partial\tilde{t}}
  \left( \frac{\partial^2\tilde\Psi}{\partial\tilde{x}^2} \right) \!\!
  & = & \!\!  
  -\frac{\partial}{\partial\tilde{x}} 
  \left( \frac{\partial^2\tilde\Theta}{\partial\tilde{x}^2} \right) -
  \hat{\gamma}\frac{\partial\tilde\Psi}{\partial\tilde{x}} \\
  \frac{\partial}{\partial\tilde{t}}
  \left( \frac{\partial^2\tilde\Theta}{\partial \tilde{x}^2} \right)\!\! 
  & = &  \!\! -\frac{\partial}{\partial \tilde{x}}
  \left( \frac{\partial^2\tilde\Psi}{\partial\tilde{x}^2} \right) - 
  \frac{1}{\hat{Ro}}\frac{\partial^2\tilde\chi}{\partial \tilde{x}^2} - 
  \hat{\gamma}\frac{\partial\tilde\Theta}{\partial\tilde{x}} \\
  \frac{\partial}{\partial\tilde{t}}
  \left(\frac{\partial^2\tilde\chi}{\partial \tilde{x}^2}\right)\!\!
  & = & \!\!  
  \frac{1}{\hat{Ro}}\frac{\partial^2\tilde\Theta}{\partial\tilde{x}^{2}} -
  \frac{1}{\hat{Ro}}\frac{\partial^2\tilde\Phi}{\partial\tilde{x}^2} -
  \hat{\gamma}\frac{\partial\tilde\chi}{\partial\tilde{x}} \\
  \frac{\partial\tilde\Phi}{\partial\tilde{t}}\ \ & = & \!\!
  \frac{\partial\tilde\Psi}{\partial\tilde{x}} - 
  \frac{1}{\hat{Ro}}\frac{\partial^2\tilde\chi}{\partial\tilde{x}^2}\, ,
\end{eqnarray}
\end{subequations}
where $(\,\tilde{\cdot}\,)$ denotes non-dimensional variables.  The
variables, $(\tilde\Psi,\,\tilde\Theta,\,\tilde\chi,\,\tilde\Phi)$,
are the non-dimensional counterparts of $(\psi_+^\prime,\,
\psi_-^\prime,\,\chi_-^\prime,\,\Phi_-^\prime)$ in
section~\ref{analysis}.

Denoting disturbances by
\[
\tilde\bfPsi\ =\ \hat{\tilde{\bfPsi}}
\exp\{i\tilde{k}\,(\tilde{x} - \tilde{c}\tilde{t})\}\, ,
\]
where $\tilde\bfPsi\ =
(\tilde\Psi,\tilde\Theta,\tilde\chi,\tilde\Phi)^{\rm T}$,
$\hat{\tilde{\bfPsi}} = (\hat{\tilde{\Psi}}, \hat{\tilde{\Theta}},
\hat{\tilde{\chi}}, \hat{\tilde{\Phi}})^{\rm T}$, and
$\tilde{c}\in\mathbb{C}$, equations~(\ref{perturbed eqns}) reduce to
\[
\tilde{\matrixbf{M}}\,\hat{\tilde\bfPsi}\ =\ \mathbf{0}\, ,
\]
 where
\[
\tilde{\matrixbf{M}}\ =\
\begin{bmatrix} 
  -\tilde{c} - \hat{\gamma}/\tilde{k}^2 & 1 & 0 & 0 \\
  1 & -\tilde{c} - \hat{\gamma}/\tilde{k}^2
  & -i/(\tilde{k}\,\hat{Ro}) & 0\\
  0 & i/(\tilde{k}\,\hat{Ro}) &
  -\tilde{c} - \hat{\gamma}/\tilde{k}^2 & -i/(\tilde{k}\,\hat{Ro}) \\
  1 & 0 & -i\, \tilde{k}/\hat{Ro} & \tilde{c}
\end{bmatrix}
.
\]\\
This leads to a normal mode solution fulfilling the fourth order
characteristic equation for $\tilde{c}$\,:
\begin{eqnarray}\label{baroclinic_freq}
  \lefteqn{ \tilde{c}^4\, +\, 
    \tilde{c}^3\left(\frac{3\hat{\gamma}}{\tilde{k}^2}\right)\, +\,
    \tilde{c}^2\left(\frac{3\hat{\gamma}^2}{\tilde{k}^4}\, -\, 
      \frac{1}{\hat{Ro}^2} - 
      \frac{1}{\tilde{k}^2 \hat{Ro}^2}-1\right)\, + } \nonumber \\
  \lefteqn{ \tilde{c}\left(\frac{\hat{\gamma}^3}{\tilde{k}^6}\, -\, 
      \frac{2\hat{\gamma}}{\tilde{k}^2 \hat{Ro}^2}\, -\, 
      \frac{\hat{\gamma}}{\tilde{k}^4 \hat{Ro}^2}\, -
      \frac{\hat{\gamma}}{\tilde{k}^2}\right)\, +\, } \nonumber \\
  \lefteqn{ \frac{1}{\hat{Ro}^2}\left(1-\frac{1}{\tilde{k}^2}\, -\, 
      \frac{\hat{\gamma}^2}{\tilde{k}^4}\right)\ \ =\ \ 0\, . } 
\end{eqnarray} 
Equation~(\ref{baroclinic_freq}) is solved numerically for varying
values of $\hat{Ro}$ and $\hat{\gamma}$.  Fig.~\ref{figA1} shows the
results for HD209458b (cf. Fig.~\ref{fig2}).

\begin{figure} 
  \vspace*{.5cm}
  \includegraphics[scale=0.42]{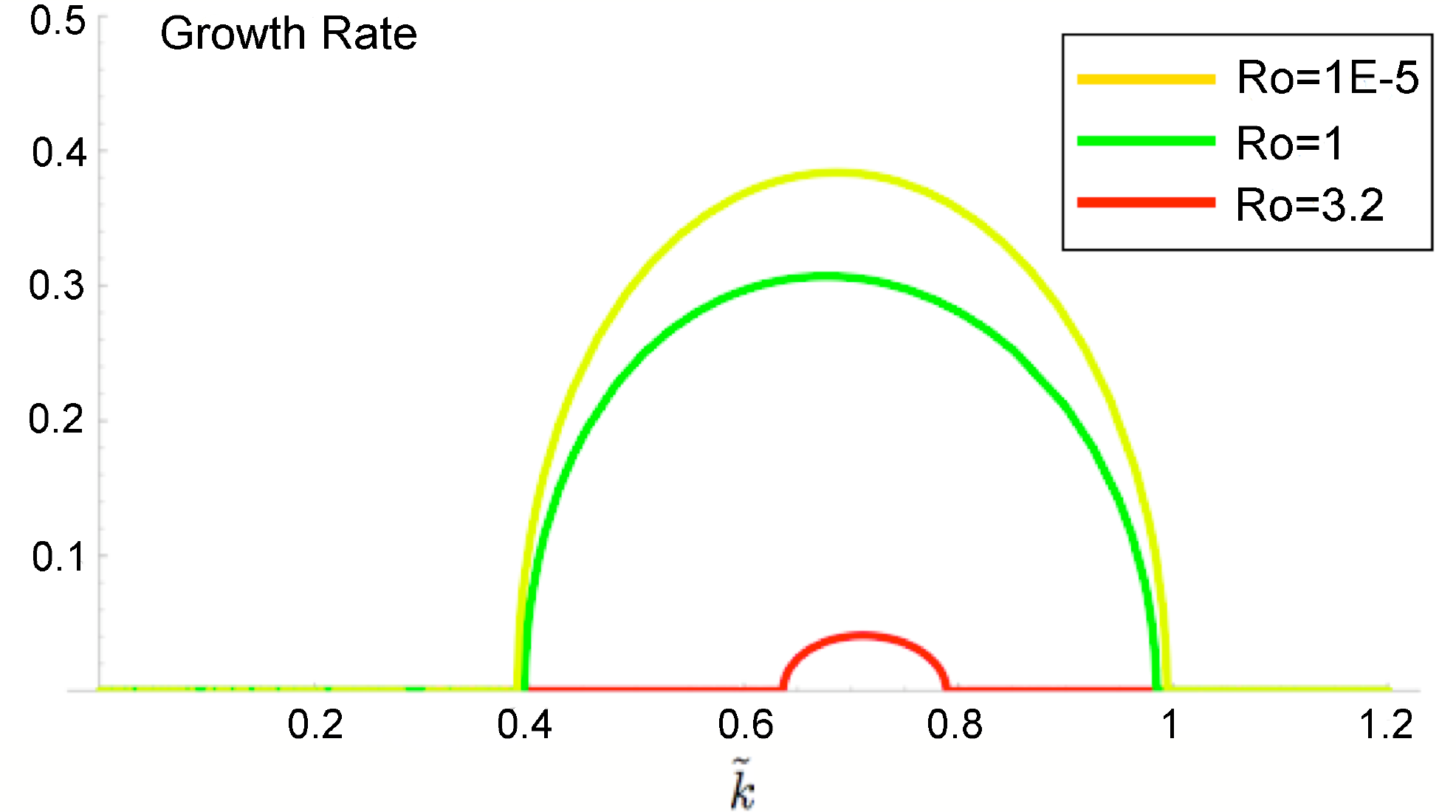}
  \vspace*{.3cm}
  \caption{Non-dimensional growth rate [$\tilde{k} \cdot
    \Im\mathfrak{m}\{\tilde{c}\}$] as a function of non-dimensional
    wavenumber $\tilde{k}$ for different values of $\hat{Ro}$. Yellow,
    green and red curves have been calculated with $\hat{Ro}$ =
    $10^{-5}$, $\hat{Ro}$ = $1$ and $\hat{Ro}$ = $3.2$,
    respectively. The value of $\hat{\gamma}$ is held constant at
    $\hat{\gamma}$ = 0.3. To obtain dimensional values, multiply the
    growth rate by $U_0 \sqrt{2}/\hat{L}_{\rm{\tiny D}}$ and
    wavenumber by $\sqrt{2}/\hat{L}_{\rm{\tiny D}}$.}
  \label{figA2}
  \vspace*{.5cm}
\end{figure}

Non-dimensional analysis is useful because it explicitly gives the
dependence of stability properties on dynamically-significant
non-dimensional numbers, such as $\hat{Ro}$ and $\hat{\gamma}$.
Extensive exploration of the solutions for a continuum of $\hat{Ro}$
and $\hat{\gamma}$ values reveals that, as $\hat{Ro}$ and/or
$\hat{\gamma}$ increase, the low wavenumber cutoff for instability
increases while the high wavenumber cutoff for instability and growth
rate decrease slightly (Fig.~\ref{figA1}). Fig.~\ref{figA2}
illustrates how the growth rates depend on $\hat{Ro}$ when
$\hat{\gamma}$ is held fixed at a typical midlatitude value for
HD209458b. As $\hat{Ro}$ increases from $\hat{Ro} << 1$ to $\hat{Ro}
\sim 1$, the growth rate decreases linearly.  However, the reduction
in growth rate is exponential as the $\hat{Ro} \sim 1$ threshold is
crossed and the two-layer linear analysis predicts stability for flows
with $\hat{Ro} > 3.3$.

To obtain the {\it dimensional} values of the growth rate from
Fig.~\ref{figA1} and Fig.~\ref{figA2}, multiply the growth rate
($\,\tilde{k}\cdot\Im\mathfrak{m}\{\tilde{c}\}\,$), for example, by
$U_0\sqrt{2}/ \hat{L}_{\rm{\tiny D}}$ and the wavenumber~($\tilde{k}$)
by $\sqrt{2}/\hat{L}_{\rm{\tiny D}}$.  The result of Fig.~\ref{figA1}
is growth rates identical to those presented in Fig.~\ref{fig2}.  The
dimensional phase speed is obtained similarly.

\label{lastpage}
\end{document}